\documentclass[a4paper,11pt]{article}

\usepackage{multicol}    
\usepackage[letterpaper, margin=1in]{geometry} 
\usepackage{algorithm}
\usepackage[noend]{algpseudocode}

\usepackage[utf8]{inputenc}
\usepackage[english]{babel}
\usepackage[T1]{fontenc}
\usepackage{amsmath}     
\usepackage[colorlinks=true]{hyperref}
\usepackage{amssymb}     
\usepackage{bm}          
\usepackage{tikz}        
\usetikzlibrary{quantikz2}
\usepackage{subfigure}   
\usepackage{url}         
\usepackage[normalem]{ulem} 
\usepackage{listings}    
\usepackage{xcolor}      
\usepackage{soul}        
\usepackage{physics}     
\pdfoutput=1
\usepackage{orcidlink}
\usepackage{subcaption}
\usepackage{float}

\renewcommand{\thefootnote}{$\dagger$}%

\usepackage{soul}
\usepackage{quantikz}

\newcommand{\tmax}{\text{max}}

\newcommand{\kb}{k_{\mathrm B}}
\newcommand{\hb}{\hbar}
\newcommand{\be}{\beta}               
\newcommand{\nB}[1]{\frac{1}{e^{\be \hb #1}-1}}

\date{}

\begin{document}

\title{\textbf{Quantum Simulation of Coupled Harmonic Oscillators: From Theory to Implementation}}

\author{
    {Viraj Dsouza}\orcidlink{0009-0007-3512-9815}$^{\text{1,a},\ast}$, 
    {Weronika Golletz}\orcidlink{0000-0002-8639-0227}$^{\text{1,b},\ast}$,
    {Dimitrios Kranas}\orcidlink{0000-0003-3035-4996}$^{\text{1,c},\ast}$,
    {Bakhao Dioum}\orcidlink{0009-0000-9588-3085}$^{\text{1,d}}$,\\
    {Vardaan Sahgal}\orcidlink{0000-0002-2293-7430}$^{\text{1,e}}$, 
    {Eden Schirman}$^{\text{2,f}}$
}

\maketitle
\def\thefootnote{\alph{footnote}}\stepcounter{footnote}\footnotetext{\href{mailto:virajdanieldsouza@gmail.com}{virajdanieldsouza@gmail.com}}
\stepcounter{footnote}\footnotetext{\href{mailto:wegolletz@gmail.com}{wegolletz@gmail.com}}
\stepcounter{footnote}\footnotetext{\href{mailto:dimitrios.kranas@gmail.com}{dimitrioskranas@gmail.com}}
\stepcounter{footnote}\footnotetext{\href{mailto:bakhao.dioum.13@gmail.com}{bakhao.dioum.13@gmail.com}} 
\stepcounter{footnote}\footnotetext{\href{mailto:vardaan.s@thewiser.org}{vardaan.s@thewiser.org}} 
\stepcounter{footnote}\footnotetext{\href{mailto:schirman.eden@gmail.com}{schirman.eden@gmail.com}} 

\def\thefootnote{\ensuremath{\ast}} 
\footnotetext{Main authors, who contributed equally to this work.}
\def\thefootnote{\arabic{footnote}} 

\def\thefootnote{\ensuremath{\dagger}} 
\def\thefootnote{\arabic{footnote}} 

\vspace{-1em} 
{\raggedright
\textsuperscript{1} The Washington Institute for STEM Entrepreneurship and Research (WISER), Washington DC, USA\\
\textsuperscript{2} Classiq Technologies, 6473104 Tel Aviv-Yafo, Israel
}
\begin{center}\today \end{center}
\vspace{1.5em} 


\begin{abstract}
  We investigate the quantum algorithm of Babbush et al. (PRX 13, 041041 (2023)) for simulating coupled harmonic oscillators, which promises exponential speedups over classical methods. Focusing on linearly connected oscillator chains, we bridge the gap between theory and implementation by developing and comparing three concrete realizations of the algorithm. First, we implement a sparse initial state preparation combined with product-formula (Suzuki–Trotter) Hamiltonian simulation. Second, we implement a fully quantum, oracle-based framework in which classical data are accessed via oracles, the Hamiltonian is block-encoded, and time evolution is performed using QSVT-based Hamiltonian simulation. Third, we propose an efficient alternative that combines the sparse state-preparation routine of the first approach with the oracle and block-encoding–based simulation pipeline of the second. We provide these implementations on Classiq, a high-level quantum design platform and provide appropriate resource benchmarks. Our simulation results show that the complex initial state preparation proposed by Babbush et al. can be circumvented at least in the linear-chain case. Finally, we illustrate two physical applications—extracting normal modes and simulating coarse-grained energy propagation—demonstrating how the algorithm connects to measurable observables. Our results clarify the resource requirements of the algorithm and provide concrete pathways toward practical quantum advantage.
\end{abstract}

\section{Introduction}

One of the most promising applications of quantum computing is the efficient simulation~\cite{Feynman1982} of systems with a high number of degrees of freedom. Qubits, the fundamental units of quantum information, are two-level quantum systems whose evolution is governed by the Schrödinger equation— a homogeneous, first-order partial differential equation that dictates unitary dynamics. This naturally raises the question of whether the evolution of physical systems governed by differential equations can be reformulated in terms of the Schrödinger equation and subsequently simulated on a quantum computer, leveraging Hamiltonian simulation techniques and appropriate classical to quantum data encoding strategies \cite{ito2023maplineardifferentialequations} to achieve quantum speed-up.

Considerable research has been devoted to establishing such mappings and demonstrating polynomial or exponential quantum speed-ups for solving differential equations. Notable advancements include a general framework for Schrödingerization of linear partial differential equations \cite{jin2023}, quantum algorithms for simulating the wave equation \cite{costa2019, Novak:2024oql}, the advection equation \cite{sato2024, Brearley_2024}, coupled harmonic oscillators \cite{babbush2023}, etc. In fault-tolerant quantum computing, these approaches hold significant promise for substantially reducing the computational cost of solving partial differential equations (PDEs) compared to classical methods. Beyond direct simulation, PDEs can also be discretized and reformulated as systems of linear equations of the form $A\ket{x}=\ket{b}$. Under certain conditions, these linear systems can be solved exponentially faster using the Harrow-Hassidim-Lloyd (HHL) algorithm \cite{harrow2009} in a fault-tolerant setting or more heuristically through variational quantum linear solvers \cite{bravo2023} on near-term quantum devices.

In this work, we investigate the quantum algorithm proposed by Babbush et al.~\cite{babbush2023} for simulating systems of coupled classical oscillators, a method that promises an exponential speed-up over its classical counterparts. 
While such theoretical advances are compelling, a significant gap persists between their abstract formulation and practical, end-to-end implementation. 
This challenge is twofold. 
First, like many quantum algorithms, its exponential advantage is contingent on access to specially constructed oracles---often treated as black boxes in theoretical analyses---with concrete, resource-efficient implementations seldom provided. 
Second, the inherent complexity of the algorithm introduces ambiguities in translating its theoretical steps into a verifiable quantum circuit. 
This disconnect not only hinders implementation but also obscures the path toward identifying the real-world applications where the demonstrated speed-up could deliver tangible value.

Our research aims to bridge this critical gap between theory and practice~\cite{babbush2025grandchallengequantumapplications}. 
We ground the abstract algorithm by focusing on a concrete physical model: the linearly connected spring-mass system. 
Using a high-level quantum software platform~\cite{classiq} to manage the underlying circuit complexity, we construct and analyze the critical components of the algorithm. 
This approach allows us to move beyond theoretical claims and quantify the precise resource requirements for a full implementation, thereby providing a grounded assessment of its feasibility on near-term and future quantum hardware. 
It is important to clarify from the outset that, as our resource analysis will confirm, harnessing this algorithm's speed-up in a practical setting is contingent upon the availability of fault-tolerant quantum devices. 
By confronting these practical hurdles head-on, this work seeks to provide a clear blueprint for translating this and similar quantum algorithms from theory into tangible application.

The primary contributions of this work are as follows:
\begin{itemize}
    \item We first validate the algorithm via a straightforward implementation that does not require complex oracle constructions. Subsequently, we develop a full end-to-end quantum circuit for a linearly connected spring-mass system, for which we contribute the explicit circuit designs of the necessary oracles.
    \item We show that for this system, the prescribed initial state preparation routine can be replaced by a more direct and resource-efficient method, leading to a notable improvement in the overall implementation.
    \item We provide a comprehensive resource analysis based on our quantum circuit implementations, comparing the results against classical numerical simulations to contextualize the costs involved.
    \item We study physical frameworks suitable for mapping to the chain of oscillators. Subsequently, we demonstrate how physical quantities of interest can be computed after executing the algorithm.
\end{itemize}

The paper is organized as follows. In Section~\ref{sec:model}, we introduce the physical model and its mapping to quantum dynamics. In Section~\ref{sec:hybrid}, we describe our hybrid classical–quantum implementation, which provides a practical validation of the algorithm. In Section~\ref{end-to-end}, we develop the full end-to-end quantum circuit framework and present a detailed resource analysis. In Section~\ref{sec:efficient-hybrid-implementation}, we propose an efficient hybrid variant that balances complexity and scalability. In Section~\ref{sec:physical_applications}, we demonstrate applications to physical problems, including vibrational spectra and coarse-grained energy transport. Finally, in Section~\ref{sec:summary}, we summarize our results and discuss implications for realizing quantum advantage in practice. Additional details of our hybrid and end-to-end implementations are provided in Appendix~\ref{AppendixA} and Appendix~\ref{AppendixB} respectively, and the scripts accompanying this work are openly available at~\cite{githubrepo}.

\section{Model}\label{sec:model}
We consider a system of \( N = 2^n \) coupled classical harmonic oscillators whose dynamics follows Newton's equations of motion:
\begin{equation}
    m_j \ddot{x}_j(t) = \sum_{k \neq j} \kappa_{jk} \left(x_k(t) - x_j(t)\right) - \kappa_{jj}x_j(t),
    \label{eq:newton}
\end{equation}
where \( m_j \) is the mass of the \( j \)-th oscillator, \( x_j(t) \) is its displacement from equilibrium, and \( \ddot{x}_j(t) \) is its acceleration. The coupling is defined by symmetric spring constants \( \kappa_{jk} = \kappa_{kj} \), and each oscillator may also be connected to a fixed wall with spring constant \( \kappa_{jj} \).

Equation~\eqref{eq:newton} can be rewritten in matrix form:  
\begin{equation}
\mathbf{M} \ddot{\vec{x}}(t) = -\mathbf{F} \vec{x}(t),
\end{equation}
where \( \vec{x}(t) \in \mathbb{R}^N \) is the displacement vector, \( \mathbf{M} \in \mathbb{R}^{N\times N}\) is a diagonal matrix with positive entries \( m_j > 0 \), and \( \mathbf{F} \in \mathbb{R}^{N\times N} \) is defined by \( f_{jj} = \sum_k \kappa_{jk} \) and \( f_{jk} = -\kappa_{jk} \) for \( j \neq k \).

The classical dynamics of the system~\eqref{eq:newton} can be mapped to quantum evolution governed by the Schr\"{o}dinger equation :
\begin{equation}
    i \frac{d}{dt} \ket{\psi(t)} = \mathbf{H} \ket{\psi(t)},
    \label{eq:schroedingereq}
\end{equation}
where the Hamiltonian \( \mathbf{H} \) in this problem is taken to be a block anti-diagonal matrix of the form:
\begin{equation}
    \mathbf{H} = -\begin{bmatrix}
        \mathbf{0} & \mathbf{B} \\
        \mathbf{B}^{\dagger} & \mathbf{0}
    \end{bmatrix}.
    \label{eq:hamiltonian}
\end{equation}
As in the original proposal~\cite{babbush2023}, the matrix \( \mathbf{B} \) is defined as:
\begin{equation}
    \sqrt{\mathbf{M}} \mathbf{B} \ket{j,k} = \begin{cases}
        \sqrt{\kappa_{jj}} \ket{j} & {\rm for}\,j=k\\
        \sqrt{\kappa_{jk}} (\ket{j} - \ket{k}) & {\rm for}\,j\neq k \\
    \end{cases}
    \label{eq:Bdefinition}
\end{equation}
acting on the basis \( \{ \ket{j,k} : j \leq k \in [N] = \{0, 1, \ldots, N-1\} \} \) of size \( \tilde{N} = N(N+1)/2 \). 
Thus, $\mathbf{B} \in \mathbb{R}^{N\times \tilde{N}}$ 
and the corresponding Hamiltonian~\eqref{eq:hamiltonian} has dimension $(N+\tilde{N}) \times (N + \tilde{N})$. 
Since quantum simulations typically require square matrices, we embed $\mathbf{B}$ into dimension $N^2\times N^2$ by zero-padding, so that $\mathbf{H} \in \mathbb{C}^{2N^2 \times 2N^2}$ and the quantum state vector $\ket{\psi(t)}$ lives in $\mathbb{C}^{2N^2}$. For more details on the mapping of classical dynamics to quantum dynamics we refer the readers to the original paper ~\cite{babbush2023} and also Appendix ~\ref{app:hybrid-2masses-initial-state} for illustration of this on a simple example.

The solution to Eq.~\eqref{eq:schroedingereq} is:
\begin{equation}
    \ket{\psi(t)} = e^{-i t \mathbf{H}} \ket{\psi(0)} = \frac{e^{-i t \mathbf{H}}}{\sqrt{2T}}
    \begin{pmatrix}
        \sqrt{\mathbf{M}} \dot{\vec{x}}(0) \\
        i \mathbf{B}^{\dagger} \sqrt{\mathbf{M}} \vec{x}(0) 
    \end{pmatrix},
    \label{eq:state-time-evolution}
\end{equation}
where \( \vec{x}(0) \) and \( \dot{\vec{x}}(0) \) are the initial positions and velocities of the oscillators, and \( T > 0 \) is the total energy:
\begin{equation}
T = E(t) + U(t),
\label{total energy}
\end{equation}
with kinetic energy 
\begin{equation}
E(t) = \frac{1}{2} \sum_j m_j \dot{x}_j(t)^2,
\end{equation}
and potential energy 
\begin{equation}
U(t)\!=\!\frac{1}{2}\!\left(\!\sum_j\! \kappa_{jj} x_j(t)^2\!+\! \sum_{k > j}\! \kappa_{jk} (x_j(t)\!-\!x_k(t))^2\!\right).
\end{equation}

In the following sections, we first provide three implementations for the above algorithm using high-level functions from the Classiq platform~\cite{classiq}.

\section{Implementation I: Sparse State Preparation and Trotterization}
\label{sec:hybrid}
To validate the quantum simulation protocol introduced in the previous section, we provide a straightforward implementation that closely mirrors the structure of the full algorithm. Our approach begins with the preparation of the initial quantum state \( \ket{\psi(0)} \), which encodes the classical configuration of the oscillator system and its initial conditions. 
To further evaluate the performance of our approach, we compare the gate complexity of the sparse state preparation to a theoretical bound for fully quantum protocols proposed in Ref.~\cite{babbush2023}. Once the initial state is prepared, we simulate the time evolution under the system Hamiltonian \( \mathbf{H} \). This Hamiltonian is first decomposed classically into a linear combination of Pauli terms and then, using the Suzuki–Trotter formula, the exponential of the sum of operators is approximated by a product of exponentials of each operator~\cite{Hatano2005Finding}. The resulting circuit operation representing $e^{-iHt}$ then acts on the initial quantum state $\ket{\psi(0)}$. This is then executed to generate the final quantum state \( \ket{\psi(t)} \). From this evolved state, the system’s dynamics can be recovered through classical post-processing.

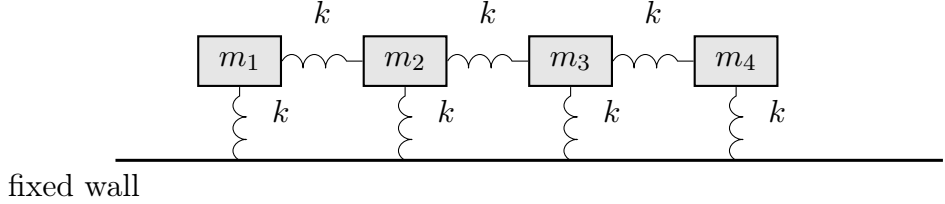
\begin{figure}[h]
\label{fig:N4figure}
\centering
\begin{tikzpicture}[scale=1.1, every node/.style={scale=1.1}]
    \draw[line width=0.4mm] (-0.5,-1.2) -- (9.5,-1.2);
    \node at (-1,-1.5) {fixed wall};

    \node[draw, thick, minimum width=1cm, minimum height=0.6cm, fill=gray!20] (m1) at (1,0) {$m_1$};
    \draw[decorate, decoration={coil,aspect=0.6,segment length=3mm}] (1,-1.2) -- (m1.south);

    \node[draw, thick, minimum width=1cm, minimum height=0.6cm, fill=gray!20] (m2) at (3,0) {$m_2$};
    \draw[decorate, decoration={coil,aspect=0.6,segment length=3mm}] (3,-1.2) -- (m2.south);

    \node[draw, thick, minimum width=1cm, minimum height=0.6cm, fill=gray!20] (m3) at (5,0) {$m_3$};
    \draw[decorate, decoration={coil,aspect=0.6,segment length=3mm}] (5,-1.2) -- (m3.south);

    \node[draw, thick, minimum width=1cm, minimum height=0.6cm, fill=gray!20] (m4) at (7,0) {$m_4$};
    \draw[decorate, decoration={coil,aspect=0.6,segment length=3mm}] (7,-1.2) -- (m4.south);

    \draw[decorate, decoration={coil,aspect=0.6,segment length=3mm}] (m1.east) -- (m2.west);
    \draw[decorate, decoration={coil,aspect=0.6,segment length=3mm}] (m2.east) -- (m3.west);
    \draw[decorate, decoration={coil,aspect=0.6,segment length=3mm}] (m3.east) -- (m4.west);

    \node at (2,0.6) {$k$};
    \node at (4,0.6) {$k$};
    \node at (6,0.6) {$k$};
    \node at (1.5,-0.6) {$k$};
    \node at (3.5,-0.6) {$k$};
    \node at (5.5,-0.6) {$k$};
    \node at (7.5,-0.6) {$k$};

\end{tikzpicture}
\caption{Four-mass oscillator system corresponding to the tridiagonal stiffness matrix \(\mathbf{F}\). 
Each mass is connected to its neighbors and to ground through springs of identical strength \(k=1\). 
Masses \(m_1\) and \(m_2\) have value \(1\); masses \(m_3\) and \(m_4\) are heavier with value \(4\).}
\label{fig:oscillator_chain}
\end{figure}

For demonstration, we consider a system of \(N\) linearly coupled classical harmonic oscillators. The mass matrix $\mathbf{M}$ is defined such that the first \(N-2\) masses have a value of $1$, while the last two masses (masses \(N-2\) and \(N-1\)) have a value of $4$, for \(N \ge 4\). For \(N < 4\), all masses are $1$. The coupling matrix $\mathbf{F}$ is a tridiagonal matrix where all diagonal elements and the first off-diagonal elements (connecting adjacent masses) are set to 1. An illustrative example for $N=4$ is provided in Figure~\ref{fig:oscillator_chain}. The initial conditions are chosen such that only the first two masses have nonzero initial displacements and velocities: \(x_0 = 0.25\), \(x_1 = -0.25\), and \(v_0 = 0.25\), \(v_1 = -0.25\). All other masses are initially in their equilibrium positions.

It is important to emphasize that this implementation approach may not offer a proven quantum advantage. Rather, it serves as a practical testbed for validating the proposed quantum algorithm. 
While the full algorithm may ultimately offer theoretical speedups, it requires quantum resources beyond the reach of current NISQ-era devices, as we will see later. 

\subsection{Initial State Preparation}\label{sec:init_state_prep} 
We prepare the initial state \(\ket{\psi(0)} \) from Eq.~\eqref{eq:state-time-evolution} using classical pre-processing followed by quantum state preparation. 
(see Appendix~\ref{app:hybrid-2masses-initial-state} for a worked example illustrating this construction for a two-oscillator system).

\begin{figure}[h!]
\centering

\begin{minipage}[t]{0.48\textwidth}
    \centering
    \includegraphics[width=\linewidth]{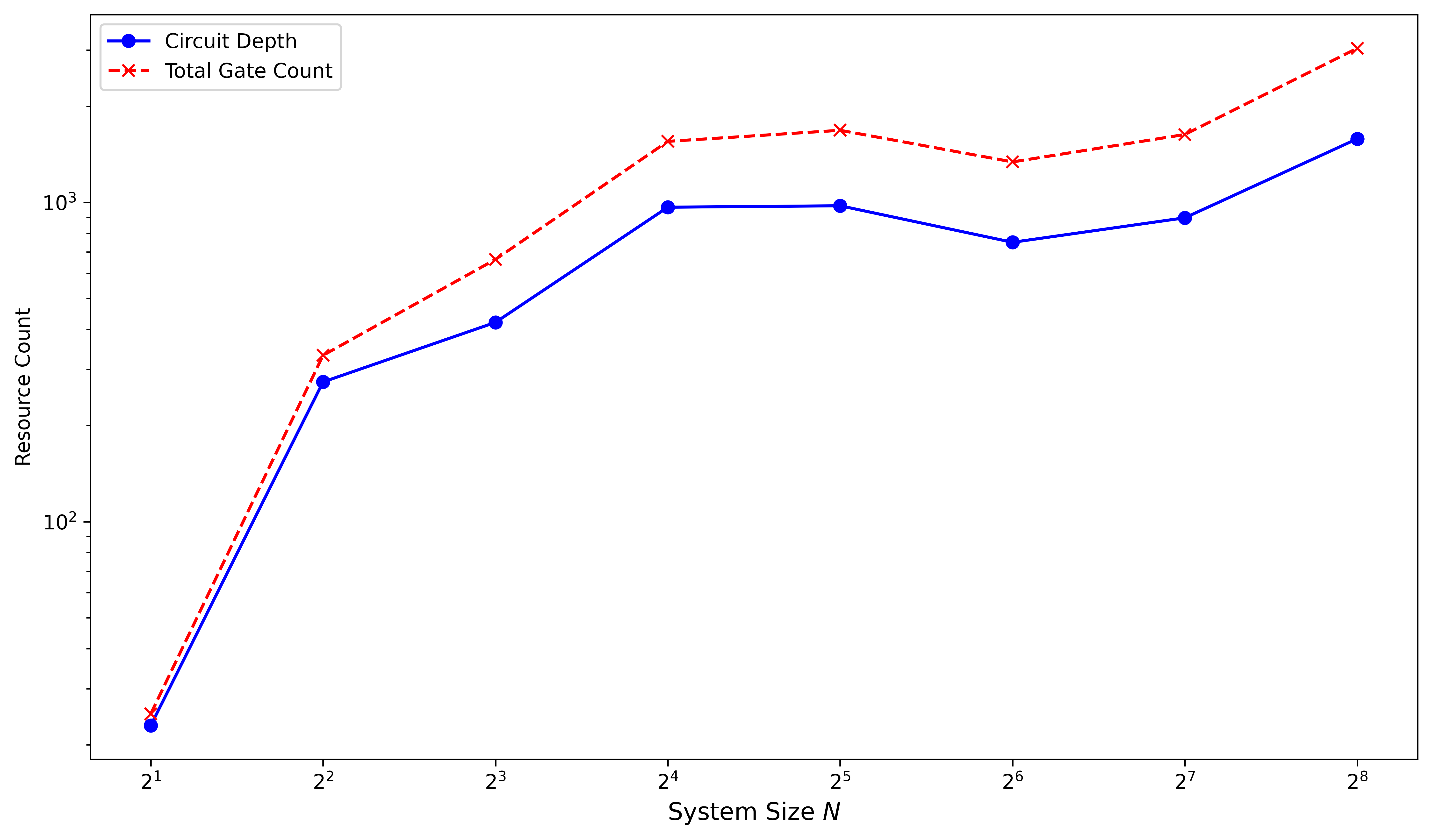}
    \caption{Circuit depth and total gate count for initial state preparation versus system size $N$.}
    \label{fig:res_vs_N}
\end{minipage}
\hfill
\begin{minipage}[t]{0.48\textwidth}
    \centering
    \includegraphics[width=\linewidth]{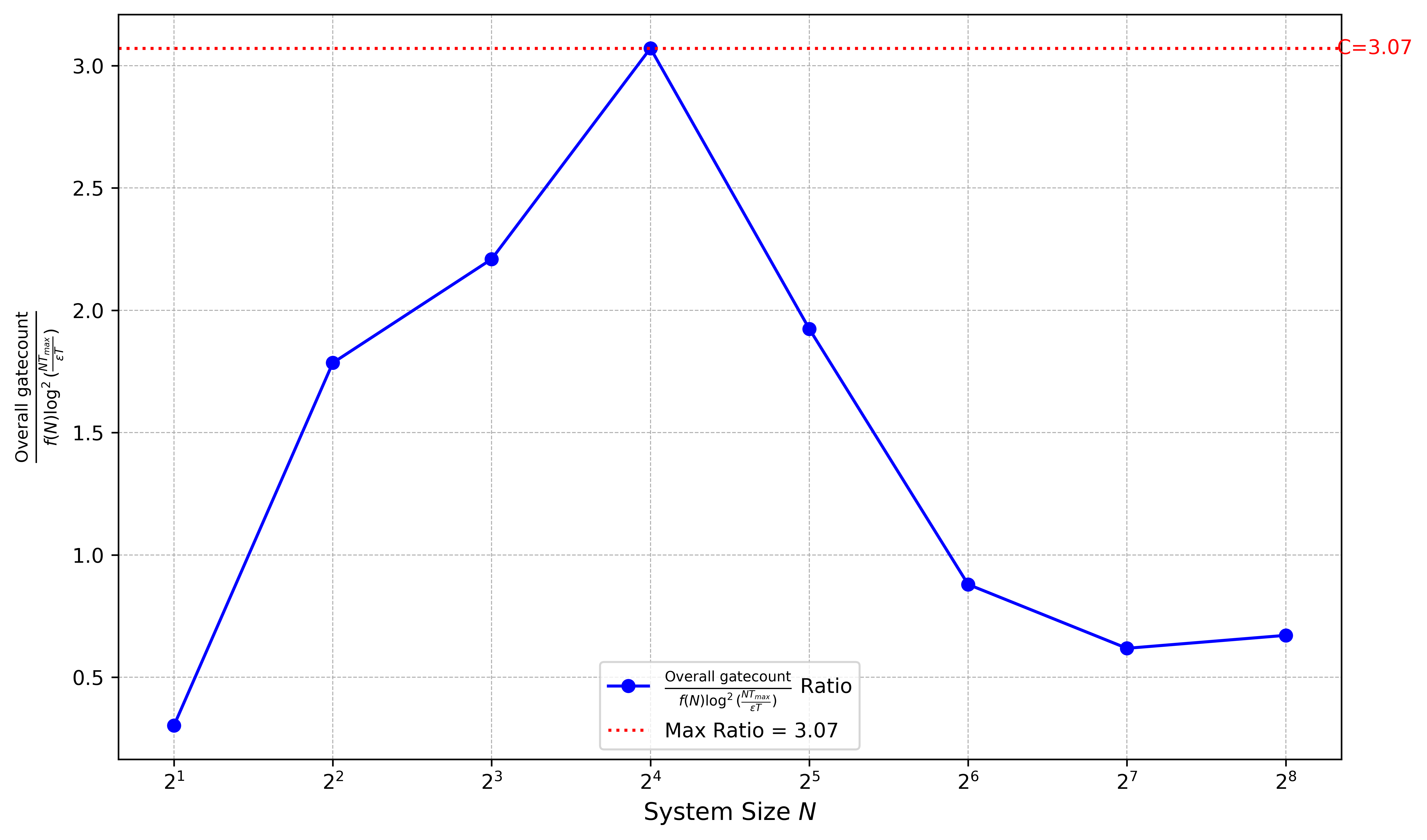}
    \caption{Ratio of experimental gate count to the scaling $f(N)\log^2(N)$, showing empirical upper bound $C=3.07$.}
    \label{fig:ratio_plot}
\end{minipage}
\label{fig:state_prep_scaling}
\end{figure}

To summarize, we classically construct the matrix \( \mathbf{B} \) from Eq.~\eqref{eq:Bdefinition} and apply it to the initial classical configuration. 
The resulting amplitudes corresponding to the kinetic and potential energy terms are loaded into a quantum register using a sparse state preparation technique ~\cite{classiq}, ~\cite{ramacciotti2023simplequantumalgorithmefficiently}. 
Figure \ref{fig:res_vs_N} shows the resources required for initial state preparation (circuit depth and total gate count\footnote{Total gate count is the sum of all single qubit unitaries and CX gates used in the quantum circuit}) as a function of system size $N$, using the fixed system configuration described in the introduction to Sec.~\ref{sec:hybrid}.

The resource counts increase rapidly for small systems before transitioning to a clear logarithmic growth for $N\approx 16$. This suggests that the cost of state preparation grows slower than any polynomial with the system size. 

\subsubsection{Gate Complexity Benchmark}
To assess the scalability of the sparse initial state preparation, we compare its gate complexity against a theoretical upper bound proposed in Ref.~\cite{babbush2023}.
In that work, a fully quantum setting is considered, where both the system parameters and initial conditions are assumed to be encoded efficiently via quantum oracles. 
Under these assumptions, the number of gates required to prepare the initial quantum state is bounded by
\begin{equation}
G_\text{in}=\mathcal{O}\left(\sqrt{d\frac{T_{\text{max}}}{T}}\log^2\left(\frac{NT_{\text{max}}}{\epsilon T}\right)\right),
\label{initial state complexity}
\end{equation}
where $\epsilon$ is the error in the state preparation, $d$ is the sparsity of the $\bm{B}$ matrix~\eqref{eq:Bdefinition}, i.e., the maximum number of nonzero entries per row.
$T$ is the total energy of the system~\eqref{total energy},
and $T_{\tmax}$ corresponds to the energy of an auxiliary system defined by the same initial conditions but with maximal mass and coupling values:
\begin{equation}
T_{\tmax}=\frac{1}{2}m_\tmax \sum_{j=1}^N\dot{x}_j^2(0)+\frac{1}{2}\kappa_\tmax \sum_{j=1}^Nx_j^2(0).
\end{equation}
where $m_{\tmax}$ and $\kappa_\tmax$ are the largest mass and coupling constant, respectively, of the original setup.
Importantly, this theoretical complexity bound~\eqref{initial state complexity} assumes that the 
$\mathbf{B}$ matrix, which encodes the spring coupling structure and contributes to the potential energy terms, is incorporated into the quantum circuit through efficient encoding.

To empirically verify the asymptotic scaling of the total gate count, we analyze its ratio with the proposed theoretical scaling function $f(N)\log^2(\frac{NT_{\text{max}}}{\epsilon T})$, where $f(N)= \sqrt{d\frac{T_{\text{max}}}{T}}$. Both $T_{max}$ and $T$ depend on system size. Throughout this analysis, we consider the linearly coupled oscillator system introduced in Sec.~\ref{sec:hybrid}. 
For this system, the sparsity of the $\mathbf{B}$ matrix is $d=2$ and we fix the maximum state preparation error to $\epsilon = 10^{-2}$.
Figure~\ref{fig:ratio_plot} illustrates this ratio for system sizes ranging from $N=2^1=2$ to $N=2^8=256$. We observe the following key characteristics:

\begin{itemize}
\item \textbf{Initial Behavior:} For smaller system sizes ($N=2$ to $N=16$), the ratio initially increases, reaching a maximum value of approximately $3.07$ at $N=16$. This initial rise reflects a \textit{pre-asymptotic regime}, where constant overheads and lower-order terms in the complexity function (which are neglected in strict Big O notation in Eq.\eqref{initial state complexity}) strongly influence the total gate count. These effects are more pronounced at small scales and can cause deviations from the idealized asymptotic behavior.
    
    \item \textbf{Asymptotic Trend:} For larger system sizes $N \geq 32$, the ratio steadily decreases. 
    This decreasing trend indicates that the gate count observed $y(N)$ grows not faster than (and in this observed range, even slower than) the proposed scaling $f(N) \log^2(N)$. 
    Thus, the total gate count is bounded by a constant multiple of this complexity for sufficiently large $N$. 
    
    Based on the results, the maximum ratio $C = 3.07$ serves as an empirical upper bound for the constant factor in the $O(f(N) \log^2(N))$ scaling. 
    The fact that the ratio does not exceed this value for larger \(N\) supports its validity as an asymptotic bound.  

\end{itemize}

It is important to note that the extension of simulations beyond $N=256$ on classical simulators proved computationally very heavy due to the rapidly increasing matrix sizes, which demands significant computational resources. Despite this limitation, the observed trend for $N \ge 32$ provides a supportive numerical evidence for the $O(f(N) \log^2(N))$ scaling of the total gate count for our quantum algorithm.


\subsection{Hamiltonian Simulation}\label{sec:ham_simulation}
To simulate the time evolution of the system governed by the Hamiltonian $ \mathbf{H} $ (Eq.~\eqref{eq:hamiltonian}), we employ a second-order Suzuki-Trotter approximation of the evolution operator $ e^{-i t \mathbf{H}} $. The first step is to decompose $ \mathbf{H} $ into a linear combination of Pauli operators, for which we use the method described in~\cite{koska2024treeapproachpaulidecompositionalgorithm}. The subsequent Trotterized simulation is performed using the Classiq platform~\cite{classiq}, which provides control over the decomposition order and the number of Trotter steps, following the framework described in~\cite{Hatano2005Finding}.

We fix the number of Trotter steps to \( r_{\text{st}} = 20 \), a value chosen to ensure that the total simulation error remains below \( 0.1 \); see Appendix~\ref{app:trotter} for a discussion of the error analysis. The evolution is performed from \( t = 0 \) to \( t = 5 \), with a discrete time step \( \Delta t = 0.1 \), using the state preparation method described  in Sec.~\ref{sec:init_state_prep}.

\begin{figure}[h] 
    \centering
        \includegraphics[width=0.4\textwidth, height=5cm]{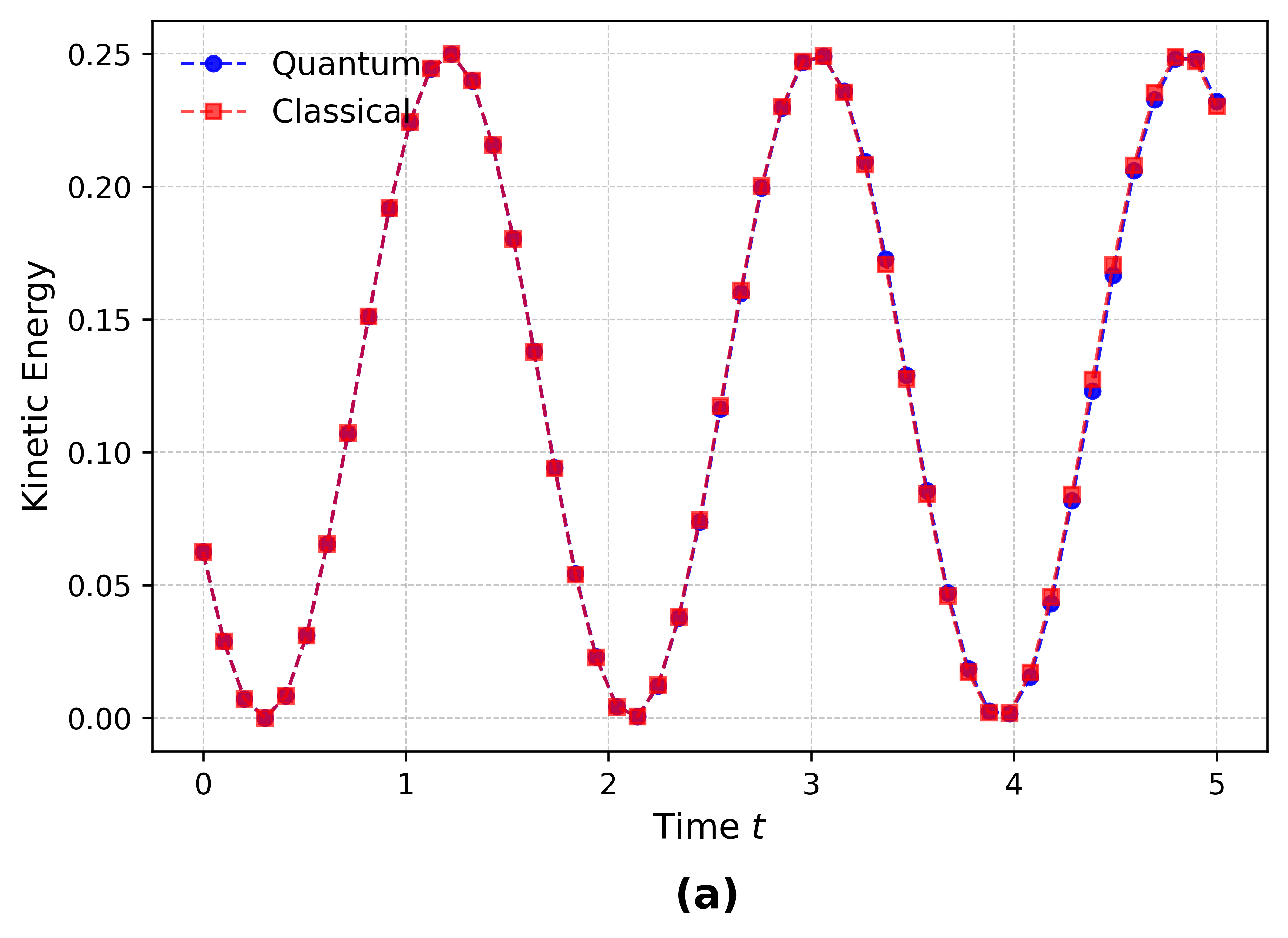}
        \label{fig:main_ke}
    \vspace{0.1cm} 
        \includegraphics[width=0.4\textwidth, height=5cm]{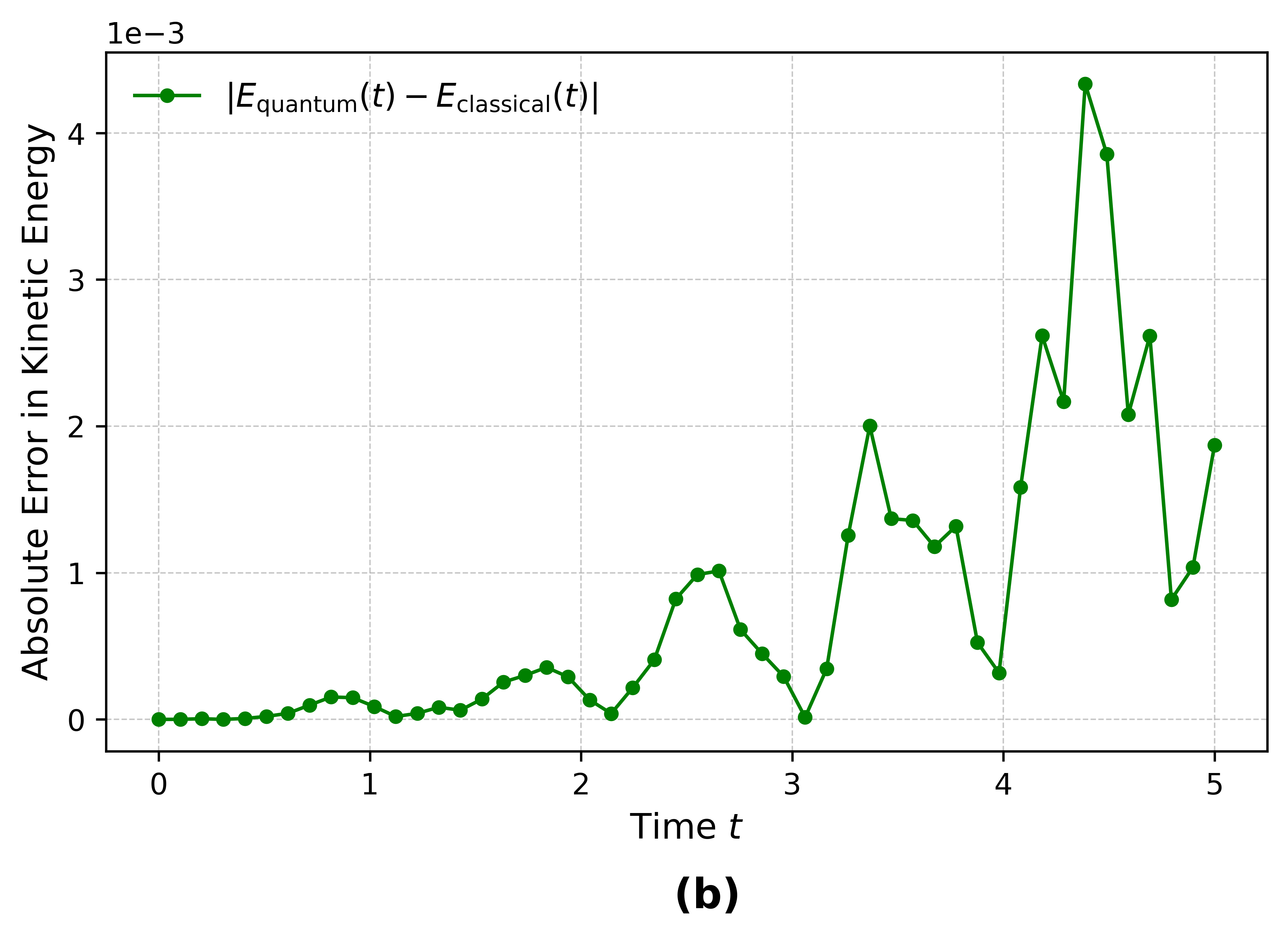}
        \label{fig:main_ke_error}
    \caption{(a) Total Kinetic energy from the quantum simulation compared with the exact classical result for \( N = 2 \) oscillators, over the interval \( t \in [0, 5] \) with time step \( \Delta t = 0.1 \). 
    (b) Absolute error between the quantum and classical results. The quantum circuit uses a second-order Suzuki–Trotter decomposition with \( r_{st} = 20 \) steps.}
    \label{fig:main_ke_combined}
\end{figure}

Figure~\ref{fig:main_ke_combined}(a) compares the total kinetic energy extracted from the quantum simulation with the exact classical result for a small system of \( N = 2 \) oscillators.
The close agreement between the two results verifies the correctness of our implementation.
The corresponding error throughout the evolution is shown in Fig.~\ref{fig:main_ke_combined}(b), demonstrating that the Trotterized quantum dynamics faithfully track the classical trajectory over time.

\subsection{Resource estimation}

\begin{figure*}
\centering
\includegraphics[width=0.95\textwidth]{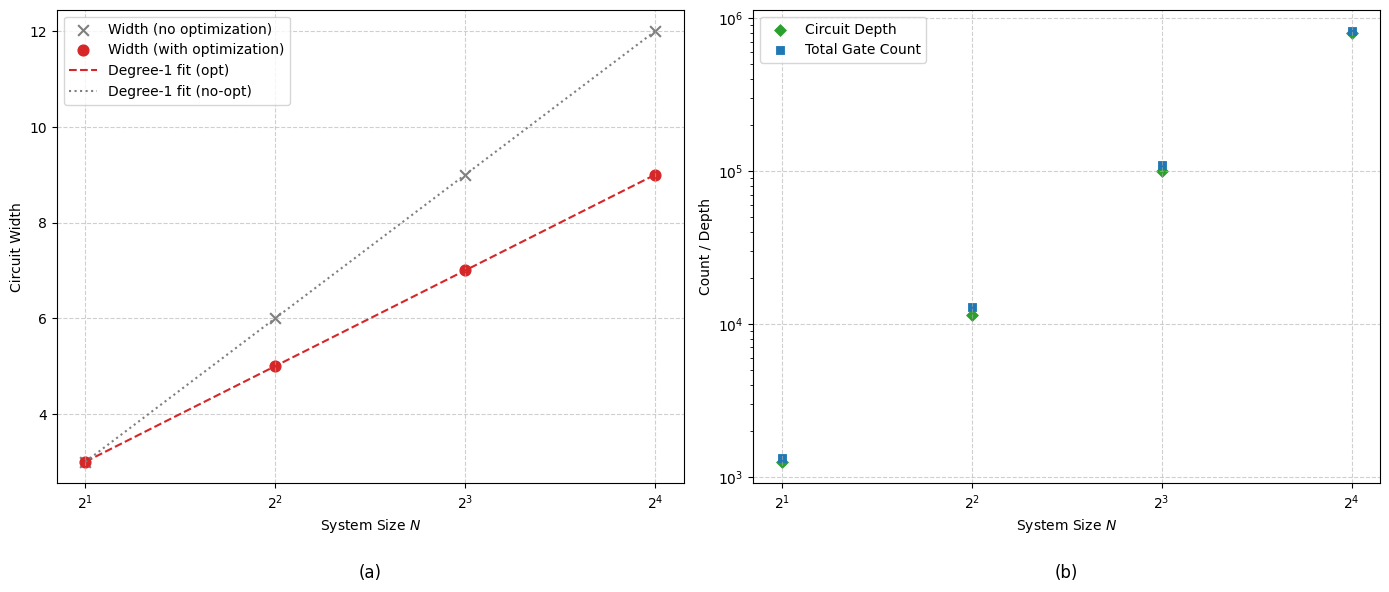}
\caption{Resource scaling and estimate for end-to-end implementation with second-order Trotterization.
(a) \textbf{Circuit Width:} Comparison of circuit width with and without synthesis optimization as a function of system size $N$.
(b) \textbf{Circuit Depth and Total Gate Count:} Circuit depth and total gate count as functions of system size $N$.}

\label{fig:Trotter_resources}
\end{figure*}

Understanding the quantum resources required is crucial for assessing the practical feasibility of quantum algorithms. In this section, we analyze the scaling of quantum resources required for the end-to-end implementation with second-order Trotterization. Figure~\ref{fig:Trotter_resources} reports the circuit width, depth, and total gate count for system sizes ranging from $N = 2$ to $N = 16$. Circuit width is shown for both the optimized and non-optimized cases. The optimization refers to the synthesis engine of Classiq automatically generating more efficient quantum circuit representations for the same high-level quantum algorithm or function. 

A first-degree polynomial fit in $\log_2(N)$ was applied to both width datasets to capture the approximate scaling trend.For the circuit depth and total gate count, no fitting was performed. The synthesis of deeper circuits becomes computationally demanding for larger $N$, and the synthesis engine could not reliably generate circuits beyond $N=16$ within reasonable classical runtime and memory limits. Since only a few data points were available, higher-degree polynomial fits (degree 2 or 3) appeared artificially accurate, indicating overfitting; these were therefore omitted.

Despite these limitations, the observed empirical trends suggest that circuit width grows linearly in $\log_2(N)$, while circuit depth and total gate count seem to also increase with the system size in a stable manner, and with a growth trend that slower than a polynomial in the system size $N$. This observed resource scaling suggests an efficiency that is nearly commensurate with the theoretical predictions outlined in Ref.~\cite{babbush2023}

\section{Implementation 2: Oracle based QSVT Framework}
\label{end-to-end}
Building on the implementation in Section~\ref{sec:hybrid}, we now turn to the construction of a fully quantum end-to-end simulation framework. While the earlier scheme leverages classical preprocessing for both state preparation and Hamiltonian decomposition, the end-to-end implementation protocol aims to replace these steps with entirely quantum procedures, thereby aligning with the structure proposed in Ref.~\cite{babbush2023}.

We construct each core building block required for the full pipeline: (i) oracle-based encoding of classical input data, (ii) quantum routines for initial state preparation, (iii) block-encoding of the system Hamiltonian, and (iv) time evolution of the block encoded Hamiltonian via Quantum Singular Value Transformation (QSVT). For each building block, we provide a detailed resource analysis to highlight the practical feasibility and potential scalability of the approach on future fault-tolerant quantum devices.

\subsection{Oracle-Based Encoding of Classical Data}\label{sec:oracles_inequality}
We use oracles to encode classical data, such as the oscillator masses 
$m_j$ and the spring constants $\kappa_{jk}$, into basis states of a quantum register.
Then, the inequality testing method is employed to map these values into quantum state amplitudes.

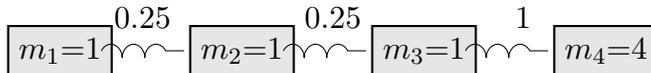
\begin{figure}[!th]
\centering
\begin{tikzpicture}[scale=1.1, every node/.style={scale=1.1}]

    \node[draw, thick, minimum width=1cm, minimum height=0.6cm, fill=gray!20] at (1.3,0) {$m_1{=}1$};

    \draw[decorate, decoration={coil,aspect=0.6,segment length=3mm}] (1.8,0) -- (2.8,0);
    \node[draw, thick, minimum width=1cm, minimum height=0.6cm, fill=gray!20] at (3.5,0) {$m_2{=}1$};

    \draw[decorate, decoration={coil,aspect=0.6,segment length=3mm}] (4.0,0) -- (5.0,0);
    \node[draw, thick, minimum width=1cm, minimum height=0.6cm, fill=gray!20] at (5.7,0) {$m_3{=}1$};

    \draw[decorate, decoration={coil,aspect=0.6,segment length=3mm}] (6.2,0) -- (7.2,0);
    \node[draw, thick, minimum width=1cm, minimum height=0.6cm, fill=gray!20] at (7.9,0) {$m_4{=}4$};

    \node at (2.3,0.4) {$0.25$};
    \node at (4.6,0.4) {$0.25$};
    \node at (6.9,0.4) {$1$};

\end{tikzpicture}
\caption{An illustrative spring mass system used for implementing the oracle based QSVT framework }
\label{fig:oscillator_chain_endtoend}
\end{figure}

The oscillator system under consideration features a spring constant matrix $\mathbf{K}$ that connects neighboring masses via weak springs of strength 0.25, except for a stronger spring of strength 1 placed between the last two masses. Except for the last block which has a mass of 4, the rest have mass of 1. An illustrative case for $N=4$ is provided in Figure~\ref{fig:oscillator_chain_endtoend}. To retrieve this classical data efficiently in a quantum algorithm, we define two data-loading oracles~\cite{babbush2023,Berry2012}:
\begin{align}
    O_{K}\ket{j,k}\ket{z} &= \ket{j,k}\ket{z \oplus \tilde{\kappa}_{jk}} = \ket{j,k}\ket{\tilde{\kappa}_{jk}}, \label{eq:oracle_K}\\
    O_{M}\ket{j}\ket{z} &= \ket{j}\ket{z \oplus \tilde{m}_j} = \ket{j}\ket{\tilde{m}_j}. \label{eq:oracle_M}
\end{align} 
where $j,k \in \{0, 1, \dots, N-1\}$ are $n=\log_2N$ sized quantum registers, $\ket{z}$ is an auxiliary quantum register initialized to $\ket{0}^{\otimes r}$, and $\tilde{\kappa}_{jk}, \tilde{m}_j$ are binary-encoded classical values. Details on the fixed-point binary representation and bit-lengths used are provided in Appendix~\ref{app:binary_representation}.

We assume the stiffness matrix 
$\mathbf{K} \in \mathbb{R}^{N\times N}$ is \emph{$d$-sparse}, meaning each row contains at most $d$ nonzero entries (including the diagonal). In a one-dimensional oscillator chain with nearest-neighbour couplings and no attachments to the wall, each interior block is connected to two blocks and boundary blocks are connected to only one block making the sparsity of $\mathbf{K}$ matrix 2. This explicit $d$-sparsity is standard in Hamiltonian-simulation oracle models and underlies QSVT-based block encodings ~\cite{Berry2012, childs2017}. To exploit the sparsity in the spring constant matrix $\mathbf{K} \in \mathbb{R}^{N\times N}$, we define an additional oracle $O_S$ that returns the column index of the $l$-th nonzero entry in row $j$:

\begin{equation}
    O_S \ket{j,l, 0^{\otimes n} } = \ket{j,l,f(j,l)},
    \label{eq:oracle_S}
\end{equation}
where $l\in \{0,1,\dots,d-1\}$ is a register of size $\lceil \log_2d \rceil$ and the function $f(j,l)$ returns the corresponding column index $k$.
With this, the combined operation to load the spring constant can be written as:

\begin{equation}
(O_K \otimes \mathbb{I}^{\otimes \lceil \log_2d \rceil})(O_S \otimes \mathbb{I}^{\otimes r})\ket{j, l, 0^{\otimes n}} \ket{0}^{\otimes r} 
= \ket{j, l, f(j, l)} \ket{\tilde{\kappa}_{j, f(j, l)}}.
\label{eq:combined_oracles}
\end{equation}

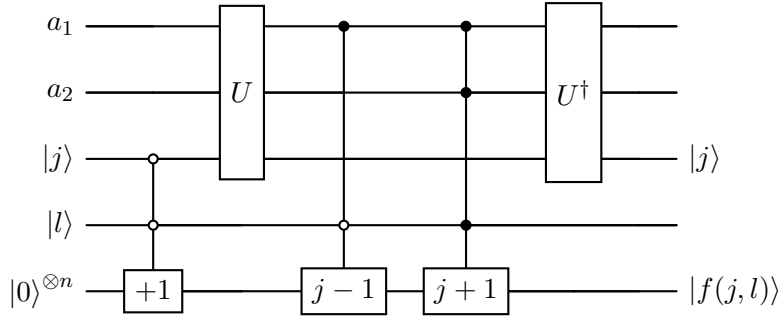
\begin{figure}[H]
\centering
\begin{quantikz}
    \lstick{$a_1$}  
    & \qw
    & \gate[3]{U} 
    & \ctrl{3} 
    & \ctrl{1} 
    & \gate[3]{U^{\dagger}} 
    & \qw
    & \qw \\
    \lstick{$a_2$}  
    & \qw  
    & \qw 
    & \qw 
    & \ctrl{2} 
    & \qw
    & \qw
    & \qw \\
    \lstick{$\ket{j}$} 
    & \octrl{1}
    & \qw
    & \qw 
    & \qw 
    & \qw
    & \qw
    & \rstick{$\ket{j}$}\\
    \lstick{$\ket{l}$} 
    & \octrl{1} 
    & \qw 
    & \octrl{1} 
    & \ctrl{1}
    & \qw
    & \qw
    & \qw \\
    \lstick{$\ket{0}^{\otimes n}$} 
    & \gate[1]{+1}
    & \qw 
    & \gate[1]{j-1}
    & \gate[1]{j+1}
    & \qw
    & \qw
    & \rstick{$\ket{f(j,l)}$} \\
\end{quantikz}
\caption{Oracle $O_S$ for the system in Figure ~\ref{fig:oscillator_chain_endtoend}}\label{fig:oracle_s_circuit}
\end{figure}

The design of the oracle $O_S$ depends on the connectivity of the oscillator network, which determines the structure of $\mathbf{K}$. 
In the case of linear connectivity, where oscillators are arranged in a 1D chain with nearest-neighbor coupling, each row of $\mathbf{K}$ has at most two nonzero entries.
The function $f(j,l)$ is then defined as:
\begin{equation}
    f(j, l) = 
    \begin{cases}
        j - 1 & \text{if } l = 0, \\
        j + 1 & \text{if } l = 1.
    \end{cases}
\end{equation}
At the boundaries, we set $f(0, 0) = 1$ (leftmost) and $f(N - 1, 0) = N - 2$ (rightmost). For all other values of $l$, we set $f(j,l)=0$. This construction works for all system sizes $N>2$. For $N=2$ the construction is trivial as the only possible value for $l$ is $0$. In this case, $f(j,l)=(j+1)\% 2$ where $\%$ denotes the remainder operation. We use high-level functional constructs from the Classiq platform to implement the quantum circuit corresponding to Eq.~\eqref{eq:combined_oracles}, with controls on $\ket{j}$ and $\ket{l}$ to reflect the sparsity structure encoded by $f(j,l)$. Figure ~\ref{fig:oracle_s_circuit} provides an overview of the circuit representing the oracle $O_S$ used in our construction. Following the action of $O_S$, the oracle $O_K$ retrieves the spring constant $\kappa_{j,f(j,l)}$ and writes its $r$-bit fixed-precision binary encoding into the target register $\ket{0}^{\otimes r}$ by conditionally applying Pauli–$X$ gates on the qubits corresponding to $1$-bits in the stored value. This implements the standard compile-time bit-oracle model for data access in sparse Hamiltonian simulation and QSVT frameworks \cite{QSVT2019, childs2017quantum}, without requiring QRAM.

\begin{figure}
\centering 
\includegraphics[width=1\textwidth]{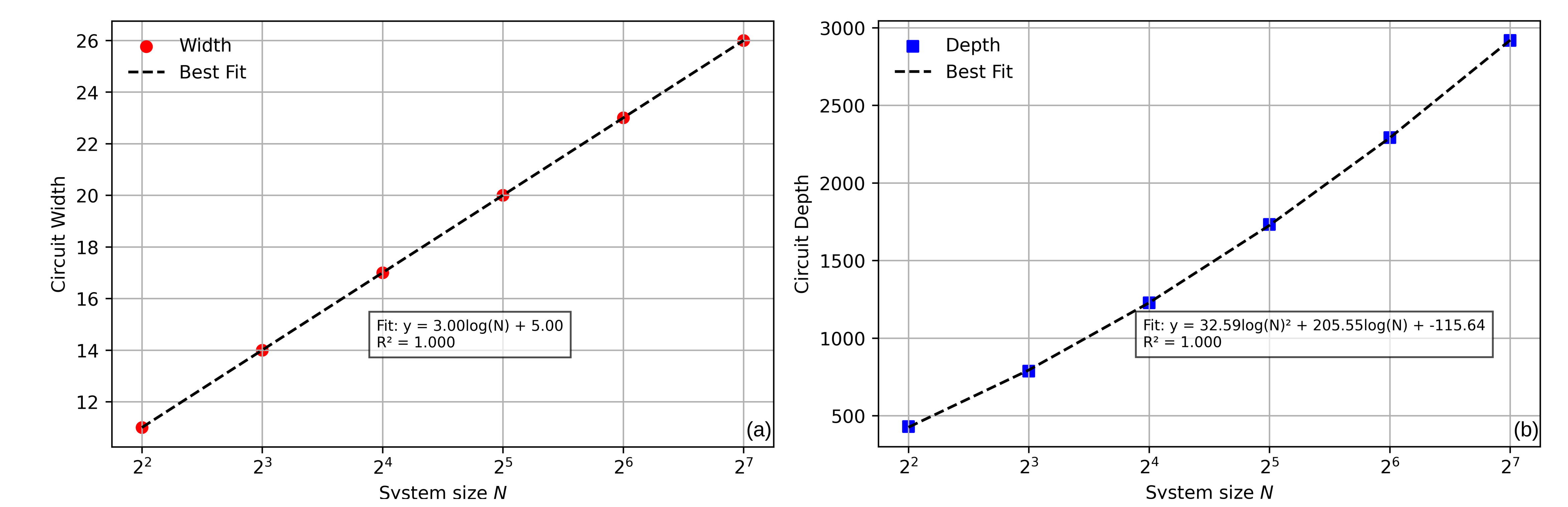} 
\caption{
Resource estimates for implementing the composed oracle that loads spring constant values in a sparse, mass-heterogeneous oscillator chain.
(a) Circuit width scales logarithmically with system size $N$, indicating efficient qubit usage.
(b) Circuit depth exhibits a logarithmic squared dependence, consistent with efficient oracle construction leveraging structured, local connectivity.
The underlying physical system consists of alternating masses (1 and 4) connected by weak springs (strength 0.25), with a stronger central spring (strength~1).
}\label{fig.qr_oracles} 
\end{figure}

Figure~\ref{fig.qr_oracles} presents quantum resource estimates—specifically, circuit width and depth—for implementing the composed oracles across various system sizes. As shown, the circuit width scales as $\mathcal{O}(\log N)$, while the depth follows $\mathcal{O}(\log^2 N)$, consistent with the expected complexity of sparse oracle construction.

\begin{figure}[H]
\centering
\begin{quantikz}
    \lstick{$a_1$}  
    & \targ{-2} 
    & \qw 
    & \qw \\
    \lstick{$a_2$}  
    & \qw \wire[u]{q} 
    & \targ{-1}
    & \qw \\
    \lstick{$\ket{j}$} 
    & \gate[1]{\begin{matrix}
        j>0 \\
        \text{Comparator}
    \end{matrix}} \wire[u]{q}
    & \gate[1]{\begin{matrix}
        j<N-1 \\
        \text{Comparator}
    \end{matrix}}\wire[u]{q}
    & \qw  \\
\end{quantikz}
\caption{The unitary $U$ of Oracle $O_S$}\label{fig:u_operation_circuit}
\end{figure}
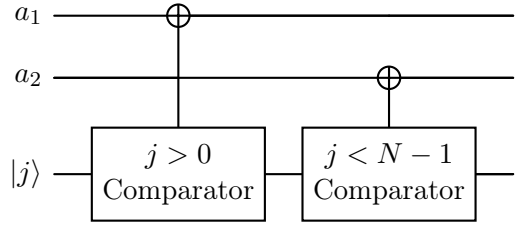

Once the classical values are encoded in basis states, i.e., $\ket{j}\ket{z} \rightarrow \ket{j}\ket{\tilde{m}_j}$, and $\ket{j,k}\ket{z} \rightarrow \ket{j,k}\ket{\tilde{\kappa}_{jk}}$, we apply the inequality testing method to map them into quantum state amplitudes~\cite{babbush2023,Sanders2019}:

\begin{equation}
    \ket{j}\ket{\tilde{\xi}_j} \rightarrow \frac{\tilde{\xi}_j}{2^{r}} \ket{j}\ket{0}^{\otimes r} + \ket{\rm junk}, 
    \label{eq:inequality_testing}
\end{equation}

where $\tilde{\xi}_j \in \{0,1,\dots,2^r-1\}$ is the integer representation of the classical value to be amplitude-encoded, such as a mass $\tilde{m}_j$ or a spring constant $\tilde{\kappa}_{jk}$, expressed using a fixed-point binary format with precision $r$ bits.
The first term contains the desired amplitude-encoded value, while the second term is an orthogonal remainder state, which can be discarded during further computation. A representative quantum circuit for this procedure is given in Figure~\ref{fig:inequality_testing_circuit}. Further details on the procedure are provided in Appendix~\ref{app:ineq_testing}.

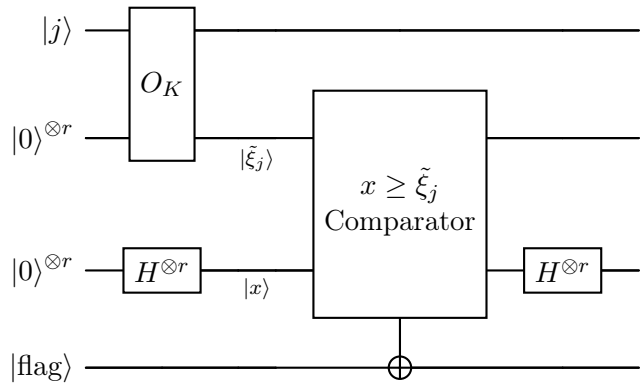
\begin{figure}[!b]
\centering
\begin{quantikz}
    \lstick{$\ket{j}$}              
    & \gate[2]{O_K} 
    & \qw   
    & \qw
    & \qw 
    & \qw 
    & \qw \\
    \lstick{$\ket{0}^{\otimes r}$}  
    & \qw  
    & \qw
    & \qw{|\tilde{\xi}_j\rangle}
    & \gate[2]{\begin{matrix}
        x \geq \tilde{\xi}_j \\
        \text{Comparator}
    \end{matrix}} 
    & \qw 
    & \qw \\
    \lstick{$\ket{0}^{\otimes r}$} 
    & \gate[1]{H^{\otimes r}}   
    & \qw
    & \qw{|x\rangle}
    & \qw \wire[d]{q} 
    & \gate[1]{H^{\otimes r}}  
    & \qw\\
    \lstick{$\ket{{\rm flag}}$} 
    & \qw  
    & \qw
    & \qw 
    & \targ{} 
    & \qw 
    & \qw 
\end{quantikz}
\caption{Inequality testing quantum circuit}\label{fig:inequality_testing_circuit}
\end{figure}
\subsection{Initial State Preparation}
Using the oracles defined in Eqs.\eqref{eq:oracle_M} and\eqref{eq:combined_oracles}, along with the inequality testing method\eqref{eq:inequality_testing}, 
we construct the initial quantum state~\eqref{eq:state-time-evolution} required for time evolution.
This state captures both the kinetic and potential energy contributions of the harmonic oscillators and is implemented on a quantum register with $2n+1$ qubits, matching the size of the padded Hamiltonian in Eq. \eqref{eq:hamiltonian}. It takes the following form~\cite{babbush2023}:
\begin{align}
\nonumber
    \ket{\psi(0)} &= \frac{1}{\sqrt{2T}}\sum_{j=1}^{N-1} \sqrt{m_j} \dot{x}_j(0) \ket{0} \ket{j} \ket{0} \\
    \nonumber
    &+ i \sum_{j=0}^{N-1} \sqrt{\kappa_{jj}} x_j(0) \ket{1} \ket{j} \ket{j} \\
    &+ i \sum_{j<k} \sqrt{\kappa_{jk}}(x_j(0) - x_k(0)) \ket{1} \ket{j} \ket{k}
    \label{eq:subnorm}
\end{align}
where $T > 0$ is the total energy~\eqref{total energy}, the indices $j\in [N]$ and $k\in [N]: \kappa_{jk}\neq 0$ being the column index of non-zero elements in the matrix $K$.

Eq.~\eqref{eq:subnorm} represents the part of the quantum state supported on a well-defined subset of qubits, while the remaining qubits—such as ancillas or intermediate registers—may contain orthogonal components that are not relevant to the target state and are therefore omitted from the expression. In practice, the amplitude of the desired component may be small due to the use of inequality testing or oracle-based loading. 
To enhance the probability of measuring the correct subspace, we employ amplitude amplification, which increases the amplitude of the target state while suppressing undesired components through Grover-like iterations. A detailed description of the amplitude amplification procedure is provided in Appendix~\ref{app:grover-iterations}.

\begin{figure}[t]
\centering 
\includegraphics[width=1\textwidth]{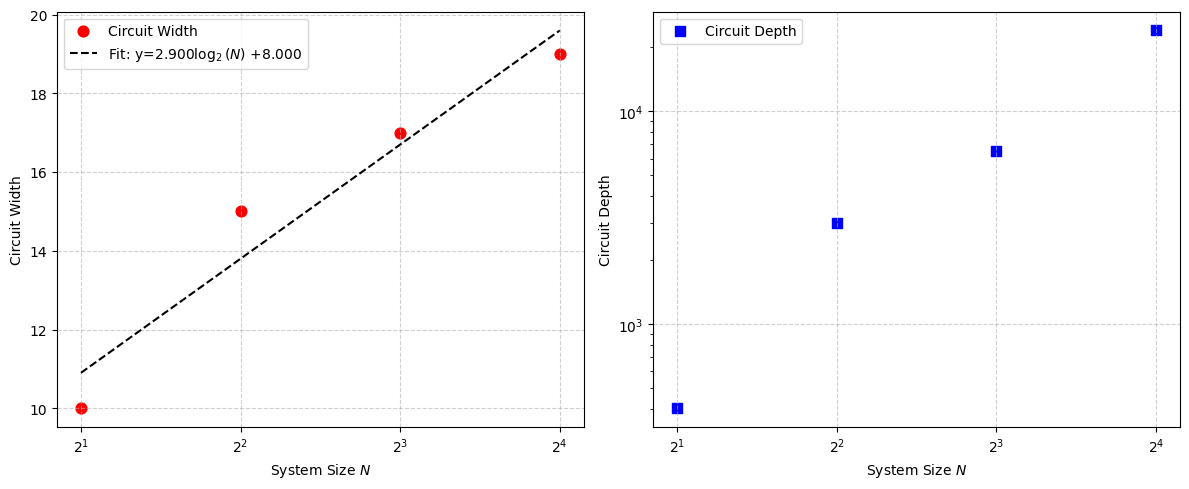} 
\caption{
Quantum circuit width (left) and depth (right) for preparing the initial state~\eqref{eq:subnorm}. The physical setup matches that of Figure~\ref{fig.qr_oracles}: A spring-mass chain with alternating masses (1 and 4), weak nearest-neighbor springs (0.25), and a stronger central spring~(1).}\label{fig:init_state_plot} 
\end{figure}

We construct the quantum circuit for preparing Eq.~\eqref{eq:subnorm}, by employing the oracles in Eqs.~\eqref{eq:oracle_M} and~\eqref{eq:combined_oracles} together with the inequality testing method~\eqref{eq:inequality_testing}. We utilize the oscillator system as described in Section 4.1. For practical implementation, specifically to keep the number of Grover iterations below 2, the initial conditions are set with the first two masses having positions \(x_0=0.25\), \(x_1=-0.25\) and velocities \(v_0=0.25\), \(v_1=-0.25\). Despite this specific choice, our implementation is general and supports arbitrary initial conditions. Implementation details are provided in Appendix~\ref{app:initial-state-preparation}. 

Figure~\ref{fig:init_state_plot} presents the width and depth of the circuit required to prepare the initial state for various sizes of the system.
The circuit width was fitted using a first-degree polynomial in $\log_2(N)$, yielding a slope of approximately $2.9 \pm 0.5$ with $R^2 = 0.94$. 
Given that only four data points were available, this result should be interpreted as \emph{suggestive rather than conclusive}—it is consistent with logarithmic growth in $N$, but additional data points across larger system sizes would be required to establish the scaling trend with statistical confidence. 
Leave-one-out analysis showed that the fitted slope varied modestly (between 2.0 and 3.5), indicating reasonable robustness of the observed trend despite the limited dataset. A corresponding fit for the circuit depth was not performed, as the available data were insufficient: reliable fits appeared only for higher-degree polynomials (quadratic or cubic) in $\log_2(N)$, which would constitute overfitting given the small sample size. 
For this reason, no polynomial model was reported for the depth dependence. Overall, these results suggest that the quantum resources scale at most polynomially with $n = \log_2(N)$, implying that the circuit for state preparation can be constructed with polynomial complexity in $n$ for spring–mass systems with local interactions.

\subsection{Block-Encoding of the Hamiltonian}

Having constructed both the data-loading oracles and the initial state entirely through quantum procedures, the next step in the end-to-end framework is to implement a block encoding of the system Hamiltonian~\eqref{eq:hamiltonian}. This encoded unitary serves as the foundational element for simulating time evolution using techniques such as QSVT.

We begin by constructing a block encoding of the matrix $\mathbf{B}^{\dagger}$, defined as
\begin{equation}
    \mathbf{B}^{\dagger} \ket{j}\! =\! \left(\sum_{j\leq k} \sqrt{\frac{\kappa_{jk}}{m_j}}\ket{j}\ket{k} \right)\! -\! \left(\sum_{j> k} \sqrt{\frac{\kappa_{jk}}{m_j}}\ket{k}\ket{j} \right),
    \label{eq:Bdaggerdefinition}
\end{equation}
which is embedded into the unitary $\mathcal{U}_{\mathbf{B}}^{\dagger}$ that captures the core structure of the Hamiltonian.

This unitary serves as the central component for a two-step embedding process. 
First, we construct a unitary whose sub-block encodes $\mathbf{B}^{\dagger}$; then, we embed this - along with its Hermitian conjugate - into a larger unitary that represents the full Hamiltonian block encoding.

We follow the general construction from Ref.~\cite{babbush2023}, but introduce two key simplifications. 
(1)~Instead of a multi-step amplitude encoding, we directly map $\ket{j, k, 0} \rightarrow \ket{j, k, a_{jk}}$, where $a_{jk} = \sqrt{\tfrac{\kappa_{jk} m_{\rm min}}{m_j \kappa_{\rm max}}}$ using the inequality testing method. (2)~We omit certain conditional gates and rotations (see steps 5–6 in Ref.~\cite{babbush2023}, Appendix \ref{app:block_encoding_B}), which are not needed under our simplified amplitude encoding scheme.
These changes reduce circuit complexity while preserving correctness in the amplitude representation, where each value is stored as an $r$-bit fixed-point binary integer. 
Implementation details are provided in Appendix~\ref{app:block_encoding_B}.

After amplitude encoding and applying a controlled-SWAP followed by a Hadamard–Z gate sequence, the resulting block-encoded state takes the form:
\begin{align}
\frac{1}{2^r\sqrt{2 d}} &\left(\sum_{j\leq k} a_{jk} \ket{j}\ket{k}\! -\! \sum_{j > k} a_{jk} \ket{k}\ket{j} \right) \ket{0}\ket{0}^{\otimes r} \ket{0} + \ket{\text{junk}}, \label{eq:block_encoding_B_dagger}
\end{align}
where $d$ is the sparsity of the matrix $\mathbf{K}$, and the $\ket{\text{junk}}$ state represents an orthogonal component outside the target ancilla subspace $\ket{0}^{\otimes r+2}$.

The state~\eqref{eq:block_encoding_B_dagger} represents the action of the unitary $\mathcal{U}_{\mathbf{B}}^{\dagger}$ on the state $\ket{j} \ket{0}^{\otimes n}\ket{0}^{\otimes r+2}$. To obtain the desired action of $\mathbf{B}^{\dagger}$, we project this on the $\ket{0}$ state of the $r+2$ ancilla qubits:
\begin{align}
    &\bra{0}^{\otimes r+2} \mathcal{U}_{\mathbf{B}}^{\dagger}\ket{j} \ket{0}^{\otimes n}\ket{0}^{\otimes r+2} \approx_{\epsilon} \frac{1}{2^r\sqrt{2d\aleph}} \mathbf{B}^{\dagger}\ket{j},
\end{align}
where $\epsilon$ is the approximation error due to inequality testing procedure and $\aleph = \kappa_{\rm max}/m_{\rm min}$.

Once the unitary $\mathcal{U}_{\mathbf{B}}^{\dagger}$ is implemented, we construct the block encoding of the full Hamiltonian block encoding by embedding both $\mathbf{B}$ and $\mathbf{B}^{\dagger}$ into a larger unitary operator. 
This construction employs two ancilla qubits to control the application of $\mathcal{U}_{\mathbf{B}}$, $\mathcal{U}_{\mathbf{B}}^{\dagger}$, and reflections about the $\ket{0}^{\otimes n}$ subspace. A step-by-step derivation is provided in Appendix~\ref{app:block_encoding_H}.

\subsection{Hamiltonian Simulation}

For Hamiltonian simulation, our primary objective is to implement the time-evolution operator, $e^{-iHt}$, as a block-encoding. To achieve this, we employ Quantum Singular Value Transformation (QSVT)~\cite{LowChuang2017},~\cite{QSVT2019}, a powerful technique capable of block-encoding a polynomial function of a matrix.

We begin by leveraging Euler's formula to decompose the time-evolution operator into its real and imaginary components: $e^{-iHt} = \cos(Ht) - i\sin(Ht)$. Since QSVT block-encodes polynomial functions, we must approximate the $\cos(Ht)$ and $\sin(Ht)$ components using polynomials. This can be accomplished through standard methods like the Jacobi-Anger expansion, which provides a series approximation for these functions.

The process then proceeds in three distinct steps:
\begin{enumerate}
    \item Block-encoding for $\cos(Ht)$: We apply QSVT to construct a block-encoding for the polynomial that approximates the cosine function, $P_{\cos}(H)$.
    \item Block-encoding for $\sin(Ht)$: A second application of QSVT generates a block-encoding for the polynomial approximating the sine function, $P_{\sin}(H)$.
    \item Linear Combination of Unitaries (LCU): The two resulting block-encodings are then combined using a linear combination of unitaries to synthesize the final block-encoding for the complete time-evolution operator, $e^{-iHt} = P_{\cos}(H) - i P_{\sin}(H)$.
\end{enumerate}

Mathematically, the QSVT procedure for a given polynomial, say for the cosine component, is constructed as a sequence of unitary transformations:

\begin{align}
    P_{\cos}(H) &= e^{i \phi_0 \sigma_z} U_H e^{i \phi_1 \sigma_z} U_H \cdots U_H e^{i \phi_k \sigma_z} \\
    \label{eq:qsvt_structure}
    e^{-iHt} &= \begin{pmatrix}
        \cos(Ht) & * \\
        * & *
    \end{pmatrix}
    - i \begin{pmatrix}
        \sin(Ht) & * \\
        * & *
    \end{pmatrix}
\end{align}

Here, $\phi_i$ are known as the QSVT angles, and $U_H$ is the block-encoding of the Hamiltonian $H$ as defined in Eq. \ref{Hamiltonian2} (refer to Appendix \ref{app:block_encoding_H} for a detailed construction of $U_H$). The phase angles $\{\phi_j\}$ are computed entirely classically using the \texttt{pyQSP}~\cite{huang2024pyqsp} python package, 
which implements numerically stable phase-finding routines for Quantum Signal Processing.

\subsection{Resource estimation}
\begin{figure*}
\centering
\includegraphics[width=0.95\textwidth]{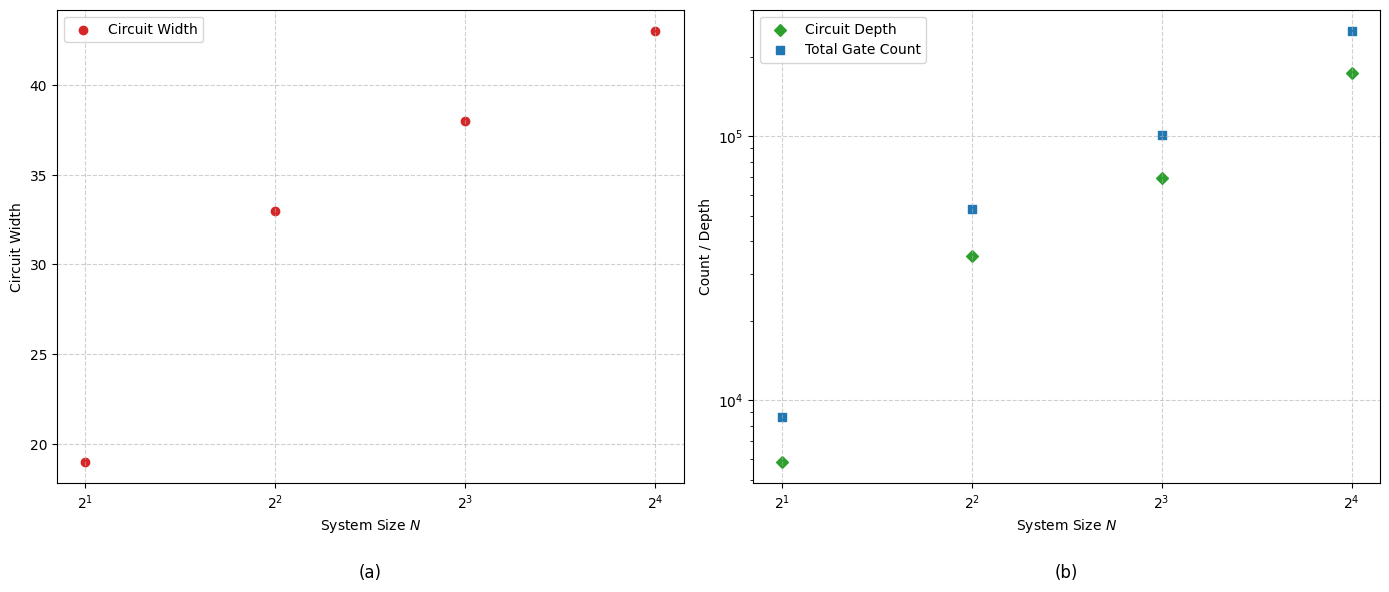}
\caption{Resource estimates for end-to-end quantum implementation.
(a) \textbf{Circuit Width:} Circuit width as a function of system size $N$. 
(b) \textbf{Circuit Depth and Total Gate Count:} Depth and total gate count as functions of system size $N$.}
\label{fig:end_to_end_quantumresources}
\end{figure*}

We present an empirical analysis of the quantum resources required to implement the end-to-end algorithm for a linearly connected spring–mass system. As shown in Figure~\ref{fig:end_to_end_quantumresources}, the circuit width, depth, and total gate count are reported for system sizes ranging from $N = 2$ to $N = 16$. 

The number of simulated data points is limited by the rapidly increasing circuit depth: for larger system sizes ($N > 16$), the synthesis engine was unable to generate the corresponding quantum circuits within practical classical computational limits. As a result, circuit construction beyond $2^4$ was not feasible on available resources. Preliminary tests indicated that reliable fits could only be obtained for higher-degree polynomials in $\log_2(N)$ (degree 2 or 3), which is not meaningful given the small dataset. Hence, we restrict our analysis to direct empirical trends observable from the plotted data. 

The results show that all three quantities—circuit width, depth, and total gate count—increase smoothly with system size, exhibiting a growth pattern consistent with at most polynomial scaling in $n = \log_2(N)$. While these trends are qualitatively in line with the expected polynomial resource dependence for systems with local interactions, additional data points across larger $N$ will be necessary to establish the precise asymptotic scaling and confirm consistency with theoretical predictions from Ref.~\cite{babbush2023}.

\section{Implementation 3: An efficient alternative implementation}
\label{sec:efficient-hybrid-implementation}

In Section 3.1.1, we established that the sparse initial state preparation method for a linearly connected spring-mass system exhibits a complexity bound consistent with a more generalized initial state preparation algorithm proposed in \cite{babbush2023}. Further, a comparison of the resource requirements shown in Figures \ref{fig:res_vs_N} and \ref{fig:init_state_plot} reveals that methodologies employing complex oracle calls and protocols like inequality testing and amplitude amplification are significantly more resource-intensive than the simpler approach detailed in our Section 3.1.

Based on these observations, we propose a more efficient end-to-end algorithm. This method combines the initial state preparation technique from Section 3.1 with the remaining components of the protocol described in Section 4. We anticipate that this integrated approach will preserve the exponential speed-up arguments presented in \cite{babbush2023}, at least for the specific case of a linearly connected spring-mass system. 

\begin{figure*}
\centering
\includegraphics[width=0.95\textwidth]{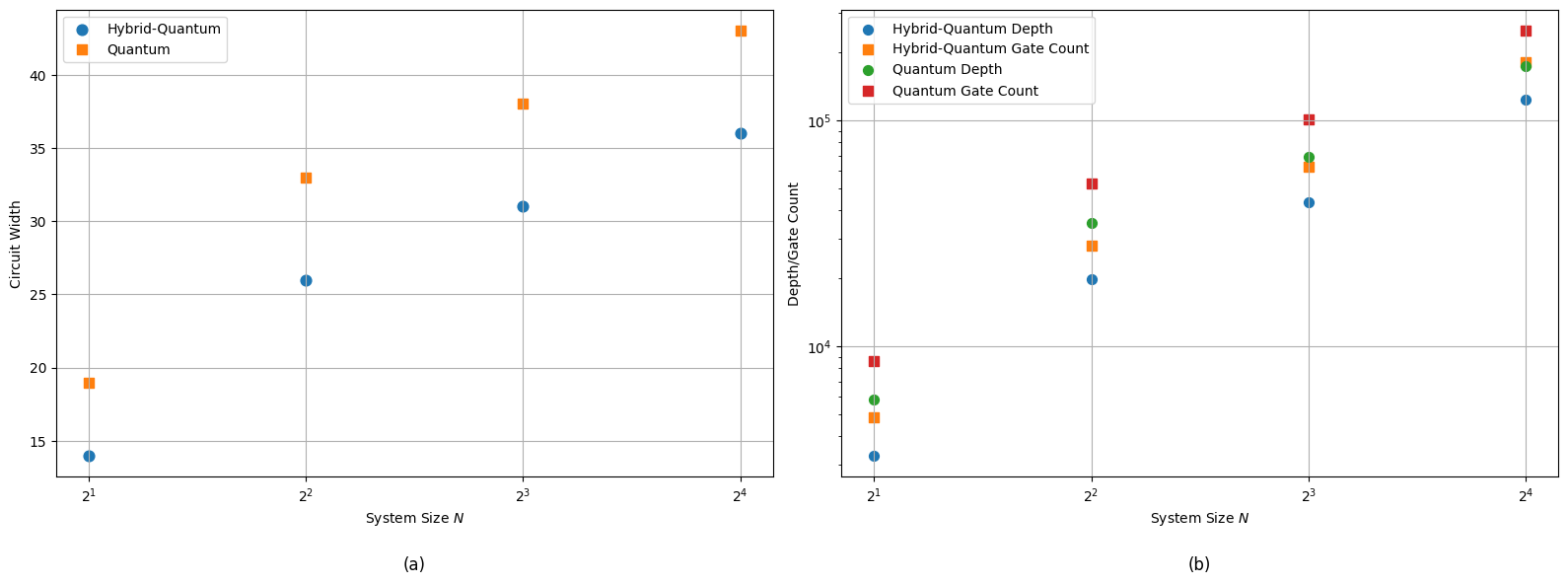}
\caption{Comparative resource scaling for end-to-end quantum and the efficient alternative implementation discussed in Section 5.
(a) Width as a function of system size $N$.
(b) Depth and Total Gate Count as a function of system size $N$.}
\label{fig:comparative_quantumresources}
\end{figure*}

Furthermore, as Figure \ref{fig:comparative_quantumresources} demonstrates, this proposed implementation requires fewer resources than the full end-to-end quantum algorithm, making it a more practical approach, particularly for linearly connected oscillators.

\section{Physical Applications}\label{sec:physical_applications}
The setup of coupled oscillators provides a widely used framework for modeling the structure and properties of various physical systems, such as lattices.
In this section, we explore how to encode such physical systems whose dynamics are governed by a Hamiltonian of the form~\eqref{total energy}, and demonstrate how to compute several physically relevant quantities from our implementation of the quantum algorithm.

In particular, our approach allows us to directly estimate the expectation values of the kinetic and potential energy, denoted by $\langle E \rangle$ and $\langle V \rangle$, respectively.
We investigate two different families of applications:
\begin{enumerate}
    \item Evaluation of the normal frequencies via the Fourier transform of the total kinetic energy $E(t)$.
    \item Computation of the total energy in subsets of neighboring oscillators.
\end{enumerate}

The first application yields the spectrum of normal frequencies in the chain of oscillators. 
Classically, this corresponds to computing the eigenvalues of the $N\times N$ matrix $\mathbf{M}^{-1}\mathbf{K}$.
While this requires diagonalizing a tridiagonal matrix, which scales as $\mathcal{O}(N^2)$ \cite{Cuppen1980, Gu1995} in the $1D$ case, the quantum algorithm extracts the normal modes directly from the time evolution of the kinetic energy, offering an exponential speedup.
The second application enables us to study the propagation of energy across different regions of the medium, thereby providing a quantum framework for modeling wave propagation.

In what follows, we discuss in detail how physical systems can be mapped to the coupled oscillator model and demonstrate how our algorithm extracts the above quantities. 

\subsection{Evaluation of normal frequencies}
The goal is to extract the set of vibrational modes of the oscillator chain directly from a quantity that can be efficiently computed from the evolved quantum state $\ket{\psi(t)}$
A natural candidate is the total kinetic energy of the system. 

This can be computed as follows. After evolving the state $\ket{\psi(t)}$ to time $t$ as described in the previous sections, one needs to apply the projector on it: 
\begin{equation}
\Pi_E=T\sum_{i=1}^N\ket{i}\bra{i}, \label{eq:K_tot_projector}
\end{equation}
which extracts the first $N$ components of the state vector~\eqref{eq:state-time-evolution}. This operation yields the total kinetic energy.
The factor of $T$ in the projector~(\ref{eq:K_tot_projector}) cancels the normalization factor $T$ that appears in the state~(\ref{eq:state-time-evolution}).

By repeating this procedure at different time instances $t$, we can construct the function $E(t)$, representing the kinetic energy as a function of time. 
For a harmonic set of oscillators, the time-averaged kinetic energy is known to be $T/2$. It is therefore convenient to define the rescaled quantity
$E'(t)=E(t)-T/2$. 

Next, we compute the discrete Fourier transform of the rescaled kinetic energy,
\begin{equation}
\tilde{E}'(\omega)=\sum_{k}E'(t_k)e^{-i\omega t_k}\Delta t,
\end{equation}
where $\{t_k\}$ is the discrete set of time samples and $\Delta t\equiv t_{k+1}-t_k$ is the timestep. The resulting spectrum $\tilde{E}'(\omega)$ exhibits sharp peaks at frequencies $2\omega_\alpha$, where ${\omega_\alpha}$ denotes the set of normal frequencies of the oscillator chain.

From the set of normal frequencies $\{\omega_\alpha\}_{\alpha=1}^N$, a series of interesting quantities with physical relevance can be computed. For instance, if the harmonic chain is weakly coupled to a large reservoir at temperature $T$, (canonical ensemble), its vibrational thermodynamics depend only on the mode frequencies. Denoting $\beta=1/(\kb T)$ and omitting any zero-frequency rigid mode, the factorised partition function is
\begin{align}
Z_{\mathrm{vib}}(\be)
  &= \prod_{\alpha=1}^{N}
     \frac{e^{-\be \hb \omega_\alpha/2}}{1-e^{-\be \hb \omega_\alpha}}.
\end{align}
The Helmholtz free energy, internal energy, entropy, and heat capacity at constant volume are
\begin{align}
\nonumber
F_{\mathrm{vib}}(T)
  &= -\kb T \ln Z_{\mathrm{vib}}
   = \frac{1}{2}\sum_{\alpha}\hb \omega_\alpha
     + \kb T \sum_{\alpha} \ln\!\bigl(1-e^{-\hb \omega_\alpha/(\kb T)}\bigr), \\
\nonumber
U_{\mathrm{vib}}(T)
  &= -\frac{\partial}{\partial \be}\ln Z_{\mathrm{vib}}
   = \sum_{\alpha}\hb \omega_\alpha\!\left(\frac12 + \nB{\omega_\alpha}\right), \\
\nonumber   
S_{\mathrm{vib}}(T)
  &= \frac{U_{\mathrm{vib}}-F_{\mathrm{vib}}}{T}
  = \kb \sum_{\alpha}\!\Big[(n_\alpha+1)\ln(n_\alpha+1)-n_\alpha\ln n_\alpha\Big], \\
\nonumber
C_V(T)
  &= \left(\frac{\partial U_{\mathrm{vib}}}{\partial T}\right)_V
   = \kb \sum_{\alpha} (\be \hb \omega_\alpha)^2 n_\alpha (n_\alpha+1),
\end{align}
where 
\begin{equation}
n_\alpha=\nB{\omega_\alpha}.
\end{equation}

\subsection{Energy Computation in Coarse-Grained Regions and the wave equation}
\begin{figure}[h]
    \centering
    \includegraphics[width=0.8\textwidth]{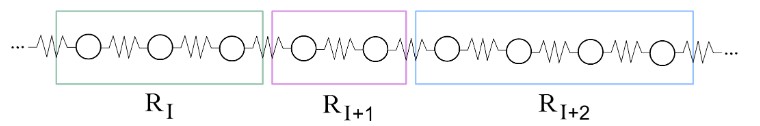}
    \caption{Illustration of the coarse-graining setup.}
    \label{Fig:coarse_grained_regions}
\end{figure}

Let is begin by deviding the system into $M$ \textit{coarse-grained regions} $R_{I=1}^M$, where each region $R_I$ contains $L$ oscillators (assumed uniform for simplicity), see Fig.~\ref{Fig:coarse_grained_regions}. This allows us to define and track local energy contributions.
\\\vspace{2pt}

\noindent\textbf{Kinetic Energy in Region $R_I$}\\
We introduce the projector
\begin{equation}
\Pi^{E}_{R_I} = \sum_{i \in R_I} \dyad{i},
\end{equation}
which acts on the first $N$ entries of $\ket{\psi(t)}$(velocity components). 
The corresponding expectation value yields the kinetic energy in region $R_I$:
\begin{equation}
E_{R_I}(t) = T \cdot \bra{\psi(t)} \Pi^{E}_{R_I} \ket{\psi(t)} = \sum_{i \in R_I} \frac{1}{2} m_i \dot{x}_i^2(t).
\end{equation}
\\\vspace{2pt}

\noindent\textbf{Potential Energy in Region $R_I$}\\
Let $\mu_l(t)$ denote the displacement of spring $l$. Define the projector
\begin{equation}
\Pi^{V}_{R_I} = \sum_{l \in \text{springs in } R_I} \dyad{N + l},
\end{equation}
which acts on the last $N(N+1)/2$ entries of $\ket{\psi(t)}$ (the spring degrees of freedom). The potential energy is then
\begin{align}
\nonumber
V_{R_I}(t) &= T \cdot \bra{\psi(t)} \Pi^{V}_{R_I} \ket{\psi(t)}=\sum_{i\in R_I}\frac{1}{2} \kappa_{ii}x_i^2(t)+\frac{1}{2}\sum_{(i,j>i) \in R_I} \frac{1}{2} \kappa_{ij} (x_i - x_j)^2(t).
\end{align}

To avoid double counting, in the last sum, we included an additional factor $1/2$ in the energy term, assigning, thus, half of the interaction energy to each of the interacting oscillators.   
\\\vspace{2pt}

\noindent\textbf{Total Energy in Region $R_I$}\\
The total energy in a region is given by
\begin{equation}
T_{R_I}(t) = E_{R_I}(t) + V_{R_I}(t).
\end{equation}

By evaluating this for each $R_I$, we obtain the \textit{coarse-grained energy profile}:
\begin{equation}
R_I \mapsto T_{R_I}(t),
\end{equation}
which describes the spatial distribution of energy across the system at time $t$.

\subsubsection{Modeling the wave equation}
This coarse-grained framework provides a natural setting to model the propagation of energy waves in the chain of oscillators. In $1+1$ dimensions, the wave equation reads
\begin{equation}
    \frac{\partial^2 T_{R_I}}{\partial t^2} = v^2 \frac{\partial^2 T_{R_I}}{\partial x^2}.
\end{equation}

Here, each region $R_I$ is regarded as an (infinitesimally) small spatial segment of the medium where energy passes through at speed $v$.  Discretizing the derivatives gives
\begin{align}
\frac{\partial T_{R_I}}{\partial t} & \approx \frac{T_{R_I}(t+\delta t)-T_{R_I}(t)}{\delta t}\\
\frac{\partial^2 T_{R_I}(t)}{\partial t^2} & \approx \frac{T_{R_{I+1}}(t)-2T_{R_{I}}(t)+T_{R_{I-1}}(t)}{(\delta t)^2},
\end{align}
where $\delta t$ is a small timestep. Similarly, one can compute the discrete spatial derivatives
\begin{align}
\frac{\partial T_{R_I}}{\partial x} & \approx \frac{T_{R_{I+1}}(t)-T_{R_I}(t)}{\delta x}\\
\frac{\partial^2 T_{R_I}(t)}{\partial x^2} & \approx \frac{T_{R_{I+1}}(t)-2T_{R_{I}}(t)+T_{R_{I-1}}(t)}{(\delta x)^2},
\end{align}
where $\delta x=aN_{R_I}$, with $a$ being the lattice spacing and $N_{R_I}$ is the number of modes in region $R_I$.

Substituting these into the wave equation allows us to evaluate the effective speed of waves
\begin{equation}
v_{I,t}=\frac{aN_{R_I}}{\delta t}\sqrt{\frac{T_{R_{I}}(t+\delta t)-2T_{R_{I}}(t)+T_{R_{I}}(t-\delta t)}{T_{R_{I+1}}(t)-2T_{R_{I}}(t)+T_{R_{I-1}}(t)}}. \label{eq:v}
\end{equation}

If $v_{I,t}$ in (\ref{eq:v}) remains constant as we vary the time and the regions in which we compute the energies, then the coarse-graining setup successfully models the wave equation.

\section{Summary and conclusion}
\label{sec:summary}
In this work, we have presented a concrete implementation of the theoretical algorithm proposed in \cite{babbush2023}. The algorithm simulates the dynamics of a linear system of $N$ coupled harmonic oscillators by mapping it onto Schr\"{o}dinger evolution. This mapping enables both state preparation and time evolution to be performed with an exponential speedup, as demonstrated in \cite{babbush2023}. Nevertheless, the original proposal is not directly executable on a quantum device. In particular, one of its key assumptions is the availability of oracles capable of efficiently encoding the model parameters (masses and spring constants) into the qubit register where the Schr\"{o}dinger dynamics unfolds. The primary objective of this work has been to construct a complete, end-to-end realization of the algorithm, with all individual components explicitly specified.

To complement our analysis, we explored a straightforward implementation, in which the evolution of the system is fully quantum, while the state preparation relies on classical data manipulation to generate the amplitudes of the quantum state. Although the theoretical speedup bound was established for the fully quantum algorithm, the first implementation in the native Classiq environment---offers an excellent framework for an initial and simpler implementation of the algorithm.  It enables early estimates of resource requirements and, for the range of system sizes $N$ accessible to circuit synthesis, was found to be less resource-demanding.

To assess our implementations, and consequently, the theoretical proposal \cite{babbush2023}, we carried out a detailed resource analysis. Using various circuit metrics, such as circuit width, depth, and gate count, we evaluated quantitatively the cost of synthesizing individual components of the algorithm as well as the full end-to-end execution. Our results provide the first evidence of the polylogarithmic scaling predicted in \cite{babbush2023}, supporting the claim of exponential speedup.  

Finally, we proposed two potential applications, where, upon sufficient scaling, the algorithm could deliver real-world benefits. First, we demonstrated how one can extract the normal modes of a system modeled as a chain of interacting oscillators. This computation leverages the quantum advantage of efficiently evaluating global properties, such as the total kinetic energy, in contrast to a classical computation of the normal frequencies, which would require storing and calculating the eigenvalues of a large matrix. Second, we proposed a protocol to map the wave equation by suitably grouping neighboring oscillators and computing the energy in each of the blocks. Following this mapping, the wave speed can be directly extracted.

In conclusion, this work constitutes a synergistic study of implementation protocols, benchmarking analysis, and the exploration of physical applications. This paves the path toward concrete, executable realizations of theoretical algorithms with potential quantum advantage, while simultaneously identifying pathways toward meaningful, real-world applications.

\section{Acknowledgments}
This work was performed as part of WISER Quantum Solutions Launchpad, and the paper was written as part of a collaboration between WISER and Classiq. The authors are listed alphabetically, and VD, WG, DK served as lead authors for this manuscript. All authors contributed to project discussions and have reviewed the final manuscript. VD would like to thank Kerem Yurtseven. The authors would also like to thank Tomer Goldfriend, Nadav Ben Ami, Amir Naveh for their support and helpful technical discussions throughout the project. 

\bibliographystyle{unsrt}
\bibliography{ref_v2} 
\appendix
\section{Appendix: Implementation I}\label{AppendixA}
This appendix provides additional technical details supporting the first implementation discussed in the main text.
We begin with a concrete example of a two-oscillator system, explicitly constructing the matrix 
(\( \mathbf{B}\)), the corresponding Hamiltonian 
\( \mathbf{H} \), and the initial quantum state 
\( \ket{\psi(0)} \).
We then include supplementary notes on the Suzuki–Trotter decomposition method used for time evolution, including numerical benchmarks of simulation error as a function of Trotter repetition count. 

\subsection{Two Masses Connected by a Spring}
\label{app:hybrid-2masses-initial-state}
To illustrate the construction of the matrix \( \mathbf{B} \), the Hamiltonian \( \mathbf{H} \), and the corresponding initial quantum state \(\ket{\psi(0)}\), we present a minimal working example with two coupled harmonic oscillators.

Let there be two masses, $m_1 = 1$ and $m_2 = 1$, connected by a single spring with coupling constant \(\kappa_{21} = \kappa_{12} = 1\), and no wall attachment springs, i.e.,  $\kappa_{11} = \kappa_{22} = 0$.

The basis for the harmonic oscillator space is
\begin{equation}
    \ket{j} = \left\{ \ket{1}, \ket{2} \right\} = 
    \left\{ \begin{pmatrix} 1 \\ 0 \end{pmatrix}, 
    \begin{pmatrix} 0 \\ 1 \end{pmatrix} \right\},
\end{equation}
and the auxiliary basis used to define the matrix \( \mathbf{B} \) is of dimension \( M = N(N+1)/2 = 3 \), and is given by
\begin{align}
    \ket{j,k} &= \left\{ \ket{1,1}, \ket{1,2}, \ket{2,2} \right\} \notag \\&=
    \left\{ \begin{pmatrix} 1 \\ 0 \\ 0 \end{pmatrix}, 
    \begin{pmatrix} 0 \\ 1 \\ 0 \end{pmatrix},
    \begin{pmatrix} 0 \\ 0 \\ 1 \end{pmatrix} \right\}.
\end{align}
The mass and force matrices are the following:
\begin{equation}
    \mathbf{M} = 
    \begin{pmatrix}
            m_1 & 0\\
            0 & m_2
            \end{pmatrix}
            =
    \begin{pmatrix}
            1 & 0\\
            0 & 1
            \end{pmatrix}
\end{equation}
\begin{equation}
    \mathbf{F} = \begin{pmatrix}
            \kappa_{11} + \kappa_{12} & -\kappa_{12}\\
            -\kappa_{21} & \kappa_{21} + \kappa_{22}
            \end{pmatrix}
            = \begin{pmatrix}
            1 & -1\\
            -1 & 1
            \end{pmatrix}
\end{equation}

To construct \( \mathbf{B} \), we apply the definition in Eq.~\eqref{eq:Bdefinition}.
For $k=j=1$:
\begin{align}
    \sqrt{\mathbf{M}}\mathbf{B}\ket{1,1} &= \sqrt{\kappa_{11}}\ket{1} \\
    \mathbf{B}\ket{1,1} &= 
    \mathbf{B}\begin{pmatrix}
        1 \\
        0 \\
        0
    \end{pmatrix} = \begin{pmatrix}
        0 \\
        0
    \end{pmatrix}
    \label{eq:ex1kj11}
\end{align}
For $k=1$ and $j=2$:
\begin{align}
    \sqrt{\mathbf{M}}\mathbf{B}\ket{1,2} &= \sqrt{\kappa_{12}}\left(\ket{1}-\ket{2}\right) \\
    \mathbf{B}\ket{1,2} &= \left(\ket{1}-\ket{2}\right) \\
    \mathbf{B}\begin{pmatrix}
        0 \\
        1 \\
        0
    \end{pmatrix} & 
    = \begin{pmatrix}
        1 \\
        -1
    \end{pmatrix}
    \label{eq:ex1kj12}
\end{align}
For $k=j=2$:
\begin{align}
    \sqrt{\mathbf{M}}\mathbf{B}\ket{2,2} &= \sqrt{\kappa_{22}}\ket{2} \\
    \mathbf{B}\ket{2,2} &=
    \mathbf{B}\begin{pmatrix}
        0 \\
        0 \\
        1
    \end{pmatrix} = \begin{pmatrix}
        0 \\
        0
    \end{pmatrix}
    \label{eq:ex1kj22}
\end{align}

Combining Eqs.~\eqref{eq:ex1kj11}–\eqref{eq:ex1kj22}, the matrix \( \mathbf{B} \) is
\begin{equation}
    \mathbf{B} = 
    \begin{pmatrix}
        0 & 1 & 0 \\
        0 & -1 & 0
    \end{pmatrix}.
    \label{eq:ex1b}
\end{equation}
It can be verified that \( \mathbf{B}\mathbf{B}^\dagger = \mathbf{F} \), consistent with the identity \( \mathbf{B}\mathbf{B}^\dagger = \sqrt{\mathbf{M}}^{-1} \mathbf{F} \sqrt{\mathbf{M}}^{-1} \), as required by definition~\cite{babbush2023}.

Following the Hamiltnian definition~\eqref{eq:hamiltonian}, we embed \( \mathbf{B} \) and \( \mathbf{B}^\dagger \) into a square \( N^2 \times N^2 \) block and construct the full Hamiltonian \( \mathbf{H} \in \mathbb{C}^{2N^2 \times 2N^2} \) as:
\begin{align}
    \mathbf{H}= \begin{pmatrix}
        0 & 0 & 0 & 0 & 0 & -1 & 0 & 0 \\
        0 & 0 & 0 & 0 & 0 & 1 & 0 & 0 \\
        0 & 0 & 0 & 0 & 0 & 0 & 0 & 0 \\
        0 & 0 & 0 & 0 & 0 & 0 & 0 & 0 \\ 
        0 & 0 & 0 & 0 & 0 & 0 & 0 & 0 \\
        -1 & 1 & 0 & 0 & 0 & 0 & 0 & 0 \\
        0 & 0 & 0 & 0 & 0 & 0 & 0 & 0 \\
        0 & 0 & 0 & 0 & 0 & 0 & 0 & 0 \\
    \end{pmatrix} 
    \label{eq:PadHex1}
\end{align}

Assuming the initial oscillator conditions are:
\begin{equation}
x_1 = 1, \quad x_2 = 2, \quad \dot{x}_1 = \dot{x}_2 = 1,
\end{equation}
the unnormalized initial quantum state \( \ket{\psi(0)} \), according to Eq.~\eqref{eq:state-time-evolution}, is
\begin{equation}
\ket{\psi(0)} = 
\begin{pmatrix}
    1 \\
    1 \\
    0 \\
    0 \\
    0 \\
    -i \\
    0 \\
    0
\end{pmatrix}.
\end{equation}
This state can be evolved under the Hamiltonian \( \mathbf{H} \) in Eq.~\eqref{eq:PadHex1} to obtain \( \ket{\psi(t)} \), describing the quantum-encoded system dynamics.

\subsection{Suzuki-Trotter Decomposition}\label{app:trotter}
This section provides additional details supporting the time evolution approach used in the main text.

\begin{figure}
\centering 
\includegraphics[width=0.5\textwidth]{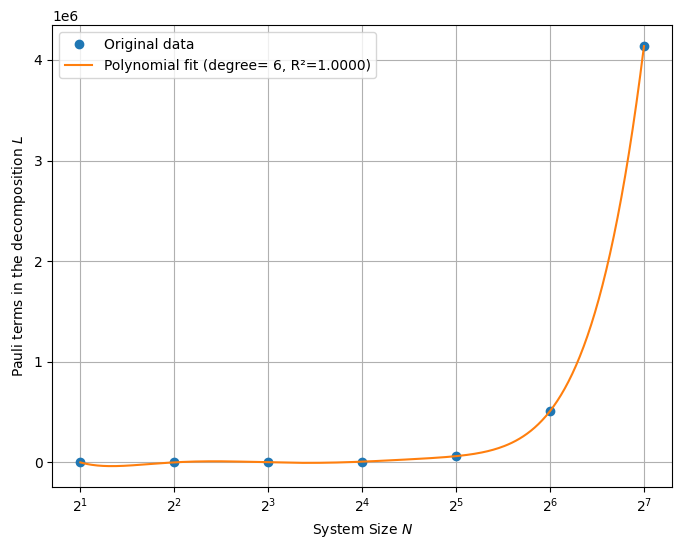} 
\caption{Empirical scaling of the number of Hamiltonian terms~\eqref{app:ham_decomposition} 
$L$ with $\log(N)$.}\label{fig:appLvslog(N)} 
\end{figure}

To simulate the unitary evolution $e^{-i\mathbf{H}t}$, we decompose the Hamiltonian $\mathbf{H}$ from Eq.~\eqref{eq:hamiltonian} as a weighted sum of normalized Hermitian operators using a Pauli basis decomposition:
\begin{equation}
    \mathbf{H} = \sum_{j=1}^L h_j H_j,
    \label{app:ham_decomposition}
\end{equation}
where each $H_j$ is a tensor product of Pauli operators (or identity), and $h_j \in \mathbb{R}$ are scalar coefficients.
Figure~\ref{fig:appLvslog(N)} shows the empirical scaling of $L$ with system size $N$ for the linearly connected system described in the introduction to Sect.~\ref{sec:hybrid}.
The observed growth is polynomial in $\log(N)$, making this decomposition efficient for practical system sizes.

While the first-order Suzuki-Trotter formula approximates time evolution operator as: 
\begin{align}
e^{-i\mathbf{H}t} &= \exp \left( -it \sum_{j=1}^{L} h_j H_j \right) \approx \left( \prod_{j=1}^{L} e^{-i h_j H_j t/ r_{\text{st}}} \right)^{r_{\text{st}}} 
\label{trotter1}
\end{align}
with error scaling as $\mathcal{O}(t^2/r_{\text{st}})$, we apply the more accurate second-order formula:
\begin{align}
&e^{-itH} = \exp \left\{ -i \left( \sum_{j=1}^{N} \frac{t_j}{2} H_j 
    + \sum_{j=N}^{1} \frac{t_j}{2} H_j \right) \right\} \approx \left( \prod_{j=1}^{L} e^{-i h_j H_j t / (2r_{\text{st}})} \prod_{j=L}^{1} e^{-i h_j H_j t / (2r_{\text{st}})} \right)^{r_{\text{st}}},
\label{eq:second-order-trotter}
\end{align}
which achieves an improved error scaling of $\mathcal{O}(t^3/r_{\text{st}})$, where $r_{\text{st}}$ is number of Trotter steps.
Higher-order Suzuki–Trotter expansions can offer even better accuracy, but at the cost of increased circuit depth and greater decomposition complexity. For our implementation, the second-order formula strikes a practical balance between accuracy and resource overhead. It is used to construct the unitary time evolution $e^{-i \mathbf{H} t}$ in the hybrid quantum simulation described in the main text.

We follow the error bound from Ref.~\cite{childs2017} for second-order formulas ($p=1$):
\begin{align}
r_{2p} = \min \{ r_{\text{st}} \in \mathbb{N} : &\frac{(2L5^{p-1}\Lambda |t|)^{2p+1}}{3r_{\text{st}}^{2p}}  \exp\left(\frac{2L5^{p-1}\Lambda |t|}{r_{\text{st}}}\right) \leq \epsilon \},
\end{align}
where $\Lambda$ is the largest absolute coefficient $h_j$, $L$ is the number of terms and $\epsilon$ is the error tolerance.

\begin{figure}
\centering 
\includegraphics[width=0.5\textwidth]{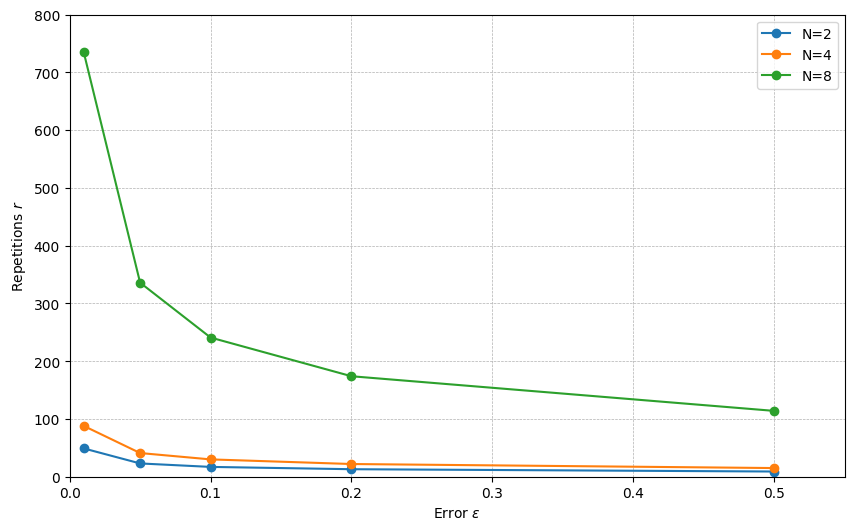} 
\caption{Required Trotter repetitions $r$ versus simulation error $\epsilon$.}\label{fig:repsvsepsilon} 
\end{figure}

Figure~\ref{fig:repsvsepsilon} shows how the required number of repetitions $r_{\text{st}}$ scales with the target error. 
Based on this analysis, we set \( r_{\text{st}} = 20 \) in the simulations presented in Section~\ref{sec:ham_simulation}, which ensures that the total simulation error remains below $0.1$ across the considered time evolution range $t\in [0,5]$.

The asymptotic gate complexity for second-order Suzuki–Trotter simulation is~\cite{childs2017}:
\begin{equation}
\mathcal{O}\left( L^{5/2} (\Lambda t)^{3/2} / \epsilon^{1/2} \right),
\end{equation}
which remains efficient under the observed scaling $L = \mathcal{O}(\text{poly}(\log N))$. While encouraging for scalability, we do not claim any formal quantum advantage in this work.

\section{Appendix: End-to-End Implementation}\label{AppendixB}
The end to end quantum algorithm introduced in \cite{babbush2023} consists of three key building blocks: initial state preparation, block encoding of the Hamiltonian, and Quantum Singular Value Transformation (QSVT) for evolving the initial state under the block-encoded Hamiltonian.
To construct these building blocks, we first introduce their essential ingredients, including oracles for accessing classical data, an inequality testing method for encoding this data into the amplitudes of the system state, and amplitude amplification to enhance the amplitude of desired quantum states while suppressing unwanted ones.

\subsection{Fixed-Point Binary Representation}\label{app:binary_representation}
Classical data are stored in matrix form: the masses are represented by a diagonal matrix $\mathbf{M}$ of size $N\times N$ with elements $m_j > 0$, and spring constants are represented by a symmetric matrix $\mathbf{K}$ of size $N \times N$ with elements $\kappa \geq 0$.
To encode them into a quantum state, we use a binary fixed-point representation, where the fractional part is represented as a binary fraction:
\begin{equation}
    m_j = m_{\rm max} [.b_{j,1}b_{j,2}\hdots]
\end{equation}
for $\forall_{j\in[N]}\, m_{\rm max} \geq m_j$ and $b_{j,i} \in \{0,1\}$.
Similarly, 
\begin{equation}
    \kappa_{jk} = \kappa_{\rm max} [.c_{jk,1}c_{jk,2}\hdots]
\end{equation}
for $\forall_{j,k\in[N]}\, \kappa_{\rm max} \geq \kappa_{jk}$ and $c_{jk,i} \in \{0,1\}$.
To balance computational efficiency and numerical accuracy,
we truncate these binary expansions using finite precision.
Let $r_m$ and $r_{\kappa}$ denote the number of fractional bits used to represent masses and spring constants, respectively.
The encoded quantities are then:
\begin{equation}
    m_j = \frac{m_{\rm max} \tilde{m}_j}{2^{r_m}}, \quad \kappa_{jk} = \frac{\kappa_{\rm max} \tilde{\kappa}_{jk}}{2^{r_{\kappa}}}
    \label{eqapp:binary-fraction}
\end{equation}
where $\tilde{m}_j$ and $\tilde{\kappa}_{jk}$ are the integer representations of the binary fractions.
The choice of $r_{m}$ and $r_{\kappa}$ governs the trade-off between precision and resource efficiency in quantum computations.

\subsection{Inequality Testing Method}\label{app:ineq_testing}
The inequality testing method enables the encoding of classical fixed-point data into quantum state amplitudes, as required in quantum algorithms such as state preparation via linear combination of amplitudes.

Assume access to an oracle that loads a fixed-point encoded classical value $\tilde{\xi}_j$ into a quantum register, such as the mass or spring constant defined in Eqs.~\eqref{eq:oracle_K}–\eqref{eq:oracle_M}:
\begin{equation}
    \ket{j}\ket{0}^{\otimes r} \xrightarrow{O_\xi} \ket{j}\ket{\tilde{\xi}_j},
\end{equation}
where $\tilde{\xi}_j \in \{0, 1, \ldots, 2^r - 1\}$ is the $r$-bit fixed-point representation of the classical value, and $O_\xi$ denotes a generic data-loading oracle.

The inequality testing method proceeds by introducing an ancilla register in uniform superposition over all $r$-bit integers, and a flag qubit initialized to $\ket{0}$:
\begin{equation}
    \ket{j}\ket{\tilde{\xi}_j} \left(\frac{1}{2^{r/2}} \sum_{x=0}^{2^r -1} \ket{x}\right)\ket{0}.
\end{equation}

A comparison operation is then performed to check whether $x \geq \tilde{\xi}_j$, and the result is encoded in the flag qubit:
\begin{equation}
    \ket{j}\ket{\tilde{\xi}_j} \frac{1}{2^{r/2}} \left( \sum_{x=0}^{\tilde{\xi}_j - 1} \ket{x} \ket{0} + \sum_{x=\tilde{\xi}_j}^{2^r - 1} \ket{x} \ket{1} \right).
\end{equation}

Applying Hadamard gates to the $r$ ancilla qubits transforms the superposition and results in an amplitude-weighted encoding that reflects the classical value $\tilde{\xi}_j$. To complete the process, the oracle used to load the value $\tilde{\xi}_j$ is uncomputed, returning the corresponding register to $\ket{0}^{\otimes r}$:
\begin{equation}
    \ket{j}\ket{0}^{\otimes r} \left( \frac{\tilde{\xi}_j}{2^r} \ket{0}^{\otimes r} \ket{0} + \ket{\rm junk} \right).
\end{equation}

The first term contains the desired amplitude-encoded version of the classical value $\tilde{\xi}_j$. The second term $\ket{\text{junk}}$ represents an orthogonal state that can be ignored in subsequent  processing.

\subsection{Initial State Preparation}\label{app:initial-state-preparation}
For initial state preparation, the methodology broadly follows Ref.~\cite{babbush2023}, in which the state $\ket{\psi(0)}$ from Eq.~\eqref{eq:state-time-evolution} is encoded using $2n+1$ qubits. This matches the dimension of the padded Hamiltonian in Eq.~\eqref{eq:hamiltonian}. The state takes the form:
\begin{align}
\nonumber
    \ket{\psi(0)} &= \frac{1}{\sqrt{2T}}\sum_j \sqrt{m_j} \dot{x}_j(0) \ket{0} \ket{j} \ket{0} + i \sum_j \sqrt{\kappa_{jj}} x_j(0) \ket{1} \ket{j} \ket{j} \notag \\&+ i \sum_{j<k} \sqrt{\kappa_{jk}}(x_j(0) - x_k(0)) \ket{1} \ket{j} \ket{k}.
    \label{eqapp:init-state-prep}
\end{align}
To construct this state efficiently, the preparation procedure is decomposed into three key components.

\paragraph{Ingredient 1: Intermediate State\\}
\noindent An intermediate superposition is constructed to encode the classical vectors $\dot{\vec{x}}(0)$ and $\vec{x}(0)$ into a quantum register, with a relative phase between the kinetic and potential energy contributions:
\begin{equation}
 \ket{\psi(0)} \mapsto \frac{1}{\sqrt{2T}}\begin{bmatrix}
    \dot{\vec{x}}(0) \\ i \vec{x}(0)
\end{bmatrix} .
\label{eqapp:ing-1}
\end{equation}
This is achieved by initializing a single qubit in the state $\ket{0}$ and applying a rotation gate $R_y(\theta)$ such that the amplitudes reflect the relative contributions of kinetic and potential energy in the system. 
The rotation angle is chosen as $\theta = 2 \cos^{-1}(a/b)$, where $a = \sqrt{m_{\rm max} \alpha}$ and $b = \sqrt{m_{\rm max} \alpha^2 + 2 \kappa_{\rm max} d \beta^2}$.
Applying this rotation yields:
\begin{equation}
    \ket{0} \mapsto \frac{1}{b} (\sqrt{m_{\rm max}}\alpha \ket{0} + \sqrt{2 \kappa_{\rm max} d}\beta \ket{1} ).
    \label{eqapp:ing-1-rot}
\end{equation}

Next, an $n$-qubit register is used to prepare quantum states encoding the normalized classical vectors.
These are loaded conditionally, controlled on the first qubit from Eq.~\eqref{eqapp:ing-1-rot}:
\begin{align}
    \ket{\dot{\vec{x}}(0)} = \frac{1}{\alpha} \sum_{j=0}^{N-1} \dot{x}_j(0) \ket{j},\,\, \ket{\vec{x}(0)} = \frac{1}{\beta}\sum_{j=0}^{N-1} x_j(0) \ket{j}.
\end{align}
Here, $\alpha$ and $\beta$ are the norms of the initial velocity vector $\dot{\vec{x}}(0)$ and the initial position vector $\vec{x}(0)$, respectively.

Finally, a $\pi/2$ phase gate is applied to the first qubit to introduce a relative complex phase, mapping the superposition to:
\begin{align}
\frac{1}{b} \left( \sqrt{m_{\rm max}} \alpha \ket{0} \ket{\dot{\vec{x}}(0)} + i \sqrt{2 \kappa_{\rm max} d} \beta \ket{1} \ket{\vec{x}(0)} \right).
\end{align}

\paragraph{Ingredient 2: Transformation of $\ket{\vec{x}(0)}$\\}
\noindent The position component of the intermediate state~\eqref{eqapp:ing-1} is transformed by applying the matrix operation $B^{\dagger} \sqrt{M}$, resulting in:
\begin{equation}
 \ket{\psi(0)} \mapsto \frac{1}{\sqrt{2T}}\begin{bmatrix}
    \dot{\vec{x}}(0) \\ i B^{\dagger} \sqrt{M} \vec{x}(0)
\end{bmatrix}.
\label{eqapp:ing-2}
\end{equation}

To achieve this, $\log(d)$ ancilla qubits are appended in the $\ket{0}$ state to enumerate the nonzero elements in each row of the matrix $\mathbf{K}$:
\begin{align}
    \ket{\vec{x}(0)}\ket{0}^{\otimes \log(d)} &\mapsto \frac{1}{\sqrt{d}\beta} \sum_{j=0}^{N-1}x_j(0)\ket{j} \sum_{l=0}^{d-1} \ket{l}. 
\end{align}

Then, the oracle $O_S$~\eqref{eq:oracle_S} is used to retrieve the column index $k$ of the $l$-th nonzero entry in row~$j$:
\begin{equation}
    \frac{1}{\sqrt{d}\beta} \sum_{j=0}^{N-1}x_j(0)\ket{j} \sum_{k\in [N]: \kappa_{jk}\neq 0} \ket{k}.
\end{equation}

Next, an oracle similar to $O_K$~\eqref{eq:oracle_K} is applied to encode $\sqrt{\kappa_{jk}/\kappa_{\rm max}}$ into the ancilla register. Using the inequality testing method (see Appendix~\ref{app:ineq_testing}), this value is encoded into the state amplitude:
\begin{align}
    \frac{1}{\sqrt{\kappa_{\rm max} d} \beta} \sum_{j,k} \sqrt{\kappa_{jk}}x_j(0)\ket{j}\ket{k}\ket{0}^{\otimes r}\ket{0} + \ket{\rm junk}, 
    \label{eqapp:ing2-after-inequality}
\end{align}
where $r$ is the bit precision, the third register corresponds to the flag qubit coming from the inequality testing method, and $\ket{\rm junk}$ is an orthogonal component that will be ignored in post-processing.

After discarding ancilla registers, the relevant part of the state is rewritten as:
\begin{align}
    \frac{1}{\sqrt{\kappa_{\rm max} d} \beta}  & \left(\sum_{j=0}^{N-1} \sqrt{\kappa_{jj}}x_j(0)\ket{j}\ket{j}\ket{0} \right.+ \sum_{j>k} \sqrt{\kappa_{jk}}x_j(0)\ket{j}\ket{k}\ket{0} + \left.\sum_{j<k} \sqrt{\kappa_{jk}}x_j(0)\ket{j}\ket{k}\ket{0} \right).
\end{align}

To enforce the ordering $j < k$ in the final state, a controlled-SWAP operation is applied to the $\ket{j}\ket{k}$ registers:
\begin{equation} \ket{j} \ket{k} \ket{0} \mapsto \begin{cases} \ket{k} \ket{j} \ket{1}, & \text{if } j > k, \\ \ket{j} \ket{k} \ket{0}, & \text{otherwise}. \end{cases} \end{equation}
This yields the transformed state:
\begin{align}
    \frac{1}{\sqrt{\kappa_{\rm max} d} \beta}  & \left(\sum_{j=0}^{N-1} \sqrt{\kappa_{jj}}x_j(0)\ket{j}\ket{j}\ket{0} \right.+ \sum_{j>k} \sqrt{\kappa_{jk}}x_j(0)\ket{k}\ket{j}\ket{1} + \left.\sum_{j<k} \sqrt{\kappa_{jk}}x_j(0)\ket{j}\ket{k}\ket{0} \right).
    \label{eqapp:controlled-swap}
\end{align}

Finally, Hadamard and Pauli-$Z$ gates ($HZ$) are applied to the ancilla qubit. This transformation maps: 
$\ket{0} \mapsto \frac{1}{\sqrt{2}} (\ket{0} + \ket{1})$ and 
$\ket{1} \mapsto \frac{1}{\sqrt{2}} (-\ket{0} + \ket{1})$,
resulting in the desired state:
\begin{align}
    &\frac{1}{\sqrt{2 \kappa_{\rm max} d} \beta}   \left(\sum_{j=0}^{N-1} \sqrt{\kappa_{jj}}x_j(0)\ket{j}\ket{j}\ket{0} \right.+ \left. \sum_{j<k} \sqrt{\kappa_{jk}}(x_j(0)-x_k(0))\ket{j}\ket{k}\ket{0} \right).
\end{align}
This completes the transformation $\ket{\vec{x}(0)} \mapsto B^{\dagger} \sqrt{M} \ket{\vec{x}(0)}$ as required in the second half of Eq.~\eqref{eqapp:init-state-prep}.

\paragraph{Ingredient 3: Transformation of $\ket{\dot{\vec{x}}(0)}$\\}
\noindent The final transformation applies $\sqrt{M}$ to the velocity component from Eq.~\eqref{eqapp:ing-2}, yielding the fully prepared initial state:
\begin{equation}
 \ket{\psi(0)} \mapsto \frac{1}{\sqrt{2E}}\begin{bmatrix}
    \sqrt{M}\dot{\vec{x}}(0) \\ i B^{\dagger} \sqrt{M} \vec{x}(0)
\end{bmatrix} .
\end{equation}

This is implemented using an oracle similar to $O_M$~\eqref{eq:oracle_M}, which encodes the values $\sqrt{m_j / m_{\rm max}}$ into an ancilla register $\ket{0}^{\otimes r}$.
Applying the inequality testing method maps these values into quantum state amplitudes:
\begin{equation}
    \ket{\vec{\dot{x}}(0)} \mapsto \frac{1}{\sqrt{m_{\rm max}}\alpha} \sum_{j=0}^{N-1} \sqrt{m_j} \dot{x}_j(0) \ket{j},
\end{equation}
where the ancilla register $\ket{0}^{\otimes r}$ and and the orthogonal component $\ket{\rm junk}$ are omitted for clarity.

Combining all three ingredients yields the final prepared quantum state:
\begin{align}
& \frac{1}{\sqrt{m_{\max} \alpha^2 + 2 \kappa_{\max} d \beta^2}} \Bigg[ \sum_{j=0}^{N-1} \sqrt{m_j} \dot{x}_j(0) \ket{0} \ket{j}\ket{0} + i \sum_{j=0}^{N-1} \sqrt{\kappa_{jj}} x_j(0) \ket{1} \ket{j}\ket{j} \notag\\
&\quad\quad\quad\,\,\,\,\,\,\,\,\,\,\,\,\,\,\,\,\,\,\,\,\,\,\,\,\,\,\,\,\,\,\,\,\,\,\,\,\,\,\,\,\quad + i \sum_{j<k} \sqrt{\kappa_{jk}} \left(x_j(0) - x_k(0)\right) \ket{1} \ket{j}\ket{k} \Bigg] + \ket{\text{junk}} \label{appeq:init_state_final}.
\end{align}

\subsection{Amplitude Amplification}\label{app:grover-iterations}
Amplitude amplification generalizes Grover’s search algorithm to boost the probability amplitude of a desired subcomponent in a quantum superposition. This technique is particularly valuable when the target state has a low initial amplitude due to probabilistic state preparation methods, such as inequality testing or oracle-based data loading. By repeatedly applying a Grover-like operator, the quantum state is rotated within a two-dimensional subspace toward the desired outcome, thereby increasing the likelihood of its successful measurement.

Any normalized quantum state $\ket{\psi}$ can be decomposed into orthogonal components associated with a "good" and a "bad" subspace:
\begin{equation}{\label{eqapp:goodbad}}
\ket{\psi}= c_1\ket{\psi_{\text{good}}}+c_2\ket{\psi_{\text{bad}}},
\end{equation}
where $\sum_{j=1}^2 |c_j|^2 = 1$.  
The objective is to amplify the amplitude of $\ket{\psi_{\rm good}}$ such that $|c_1| \rightarrow 1$, while $|c_2| \rightarrow 0$.

This is accomplished using the Grover operator: 
\begin{equation} \label{eqapp:grover}
 \mathbf{Q}= - \mathbf{A} \mathbf{S}_{0} \mathbf{A}^{\dagger} \mathbf{S}_{\ket{\psi_{\text{good}}}},
\end{equation} 
where $\mathbf{A}$ is the state preparation unitary such that $\mathbf{A}\ket{0} = \ket{\psi}$, and $\mathbf{S}_{0}=\mathbb{I}-2\ket{0}\bra{0}$ and $\mathbf{S}_{\ket{\psi_{\text{good}}}}= \mathbb{I}-2\ket{\psi_{\text{good}}}\bra{\psi_{\text{good}}}$ are reflections about the zero state and the target state, respectively. 
The operator $\mathbf{Q}$ acts as a rotation by an angle $2\theta$ in the subspace spanned by $\ket{\psi_{\rm good}}$ and $\ket{\psi_{\rm bad}}$, where the initial amplitude is given by $c_1= \sin{\theta}$ and $c_2= \cos{\theta}$.
The state $\ket{\psi}$ after $w$ applications of $\mathbf{Q}$ evolves as:
\begin{equation}
\begin{split}
\mathbf{Q}^w\mathbf{A}\ket{0}&= \sin{\left[(2k+1)\theta\right]} \ket{\psi_{\text{good}}} \\&+\cos{[(2k+1)\theta]}  \ket{\psi_{\text{bad}}}.
\end{split}
\end{equation}
The amplitude of $\ket{\psi_{\rm good}}$ is maximized by choosing $w= \lfloor\frac{\pi}{4\theta}\rfloor$.

This formulation allows systematically amplifying the probability of success in measuring the desired state, even when it is initially suppressed due to low overlap with the total quantum state. Construction of the $\mathbf{S}_{\ket{\psi_{\text{good}}}}$ reflection operator depends on the structure of the state and typically involves marking the target configuration using phase flips conditioned on specific register values.

For the prepared state in Eq.\eqref{appeq:init_state_final}, the component labeled $\ket{\text{junk}}$ corresponds to the undesired subspace $c_2\ket{\psi_{\text{bad}}}$, while remaining terms represent the desired component $c_1 \ket{\psi_{\text{good}}}$. 
In systems with $\kappa_{jj}=0$, the second term in~\eqref{appeq:init_state_final} vanishes, simplifying the target state. 
With this, the full state can be expressed in the decomposition of Eq.~\eqref{appeq:init_state_final} as: 
\begin{align}
&c_1 \Bigg[\frac{1}{c_1\sqrt{m_{\max} \alpha^2 + 2 \kappa_{\max} d \beta^2}} \times \notag\\ &\,\,\,\,\,\,\,\,\Bigg( \sum_{j=0}^{N-1} \sqrt{m_j} \dot{x}_j(0) \ket{0} \ket{j}\ket{0} + i \sum_{j<k} \sqrt{\kappa_{jk}} \left(x_j(0) - x_k(0)\right) \ket{1} \ket{j}\ket{k} \Bigg)\Bigg] 
+c_2 \Bigg[\frac{1}{c_2}\ket{\text{junk}}\Bigg] ,\label{eqapp:ampaa}
\end{align}

To perform amplitude amplification on this state, it is necessary to implement the reflection operator $\mathbf{S}_{\ket{\psi_{\text{good}}}}$ 
which flips the phase of the desired component while leaving the orthogonal part unaffected. In this implementation, the operator is realized by applying a $Z$ gate to the first qubit conditioned on all $2r+1$ ancilla qubits being in the $\ket{0}$ state and the logical condition $j=k-1$. For the special case $k=0$, an $X-Z-X$ gate sequence is applied to the first qubit to perform a controlled phase flip.

The reflection $\mathbf{S}_0$ is constructed using standard multi-qubit control techniques.
All qubits are first inverted with $X$, followed by a multi-controlled $Z$ gate, where one qubit is designated as the target and the rest act as controls.
A second round of $X$ gates restores the original state basis.
This construction ensures a phase flip only when the full system register is in the $\ket{0}$ state, thereby completing the amplitude amplification procedure.

\subsection{Block encoding of \textbf{$\mathbf{B}^{\dagger}$}}\label{app:block_encoding_B}
The block encoding of the matrix $\mathbf{B}^{\dagger}$~\eqref{eq:Bdaggerdefinition}
is realized by constructing a unitary operator $\mathcal{U}_{\mathbf{B}}^{\dagger}$.

The procedure begins by preparing a quantum register in the state $\ket{j}$. A second register is then introduced to enumerate the nonzero elements in each row of the stiffness matrix $\mathbf{K}$, producing the uniform superposition:
\begin{equation}
    \ket{j}\ket{0}^{\otimes n} \mapsto \frac{1}{\sqrt{d}} \ket{j}\sum_{l=0}^{d} \ket{l},
\end{equation}
where $d$ is the sparsity of the matrix $\mathbf{K}$.
Applying the oracle $O_S$~\eqref{eq:oracle_S} yields the corresponding column indices $k$ for which $\kappa_{jk} \ne 0$:
\begin{equation}
    \frac{1}{\sqrt{d}} \ket{j}\sum_{k\in[N]:\kappa_{jk}\neq 0} \ket{k}.
\end{equation}
Next, the amplitude $a_{jk} = \sqrt{\frac{\kappa_{jk}}{m_j \aleph}}$ is encoded using an oracle analogous to $O_K$~\eqref{eq:oracle_K}, where $\aleph = \kappa_{\rm max} / m_{\rm min}$. This value is loaded into an ancilla register of $r$ qubits initialized to $\ket{0}^{\otimes r}$, resulting in the transformation:
\begin{equation}
    \frac{1}{\sqrt{d}} \ket{j}\sum_{k} \ket{k}\ket{0}^{\otimes r} \mapsto \frac{1}{2^r\sqrt{d}} \ket{j}\sum_{k} \ket{k}\ket{a_{jk}}.
\end{equation}
To incorporate the amplitude into the quantum state, the inequality testing procedure (see Sec.~\ref{sec:oracles_inequality}) is applied, yielding:
\begin{equation}
    \frac{1}{2^r\sqrt{d}} \sum_k a_{jk} \ket{j}\ket{k}\ket{0}^{\otimes r}.
\end{equation}
An additional qubit is appended, and a controlled-SWAP operation~\eqref{eqapp:controlled-swap} is applied to distinguish the $j<k$ and $j>k$ branches, producing:
\begin{align}
    \frac{1}{2^r\sqrt{d}} &\Big(\sum_{j\leq k} a_{jk} \ket{j}\ket{k} \ket{0}+ \sum_{j > k} a_{jk} \ket{k}\ket{j} \ket{1}\Big) \ket{0}^{\otimes r}.
\end{align}
Finally, a Hadamard and $Z$ gates are applied to the swap flag qubit.
This transforms the state into:
\begin{align}
    \frac{1}{2^r\sqrt{2 d}} &\Big(\sum_{j\leq k} a_{jk} \ket{j}\ket{k}- \sum_{j > k} a_{jk} \ket{k}\ket{j} \Big) \ket{0}^{\otimes r} \ket{0}.
\end{align}
Here, any orthogonal remainder outside the target ancilla subspace is omitted. This completes the construction of the block-encoding unitary $\mathcal{U}_{\mathbf{B}}^{\dagger}$. For obtaining the block encoding of $\mathbf{B}$ we take the inverse of the unitary $\mathcal{U}_{\mathbf{B}}^{\dagger}$. 

\subsection{Block encoding of the Hamiltonian}\label{app:block_encoding_H}
\begin{figure}[!b]
\centering
\begin{quantikz}
    \lstick{$\ket{0}_{\mathbf{P}}$} & \gate{H} & \qw & \ctrl{1} & \ctrl{1} & \gate{H} & \qw \\
    \lstick{$\ket{0}_{\mathbf{U}}$} & \gate[2]{\text{State prep}} & \ctrl{1} & \ctrl{1} & \octrl{1} & \octrl{1} & \qw \\
    \lstick{$\ket{\Phi}_{\mathbf{T}}$} & \qw & \gate{-U_B} & \gate{2P-I} & \gate{2P-I} & \gate[1]{-U_B^\dagger} & \qw 
\end{quantikz}
\caption{Circuit performing the block encoding of the Hamiltonian. Register sizes: $\ket{0}_{\mathbf{P}}$ has \(1\) qubit, $\ket{0}_{\mathbf{U}}$ has \(1\) qubit, and the target $\ket{\Phi}_{\mathbf{T}}$ has \(2n + r + 2\) qubits.}\label{fig:BEH_circuit}
\end{figure}
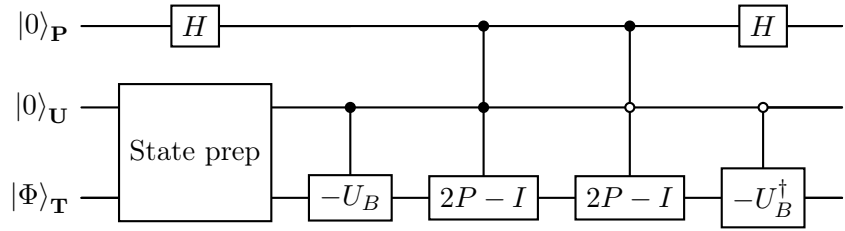

Given the block encoding $\mathcal{U}_{\mathbf{B}}$ of the matrix $\mathbf{B}$, the Hamiltonian~\eqref{eq:hamiltonian} can be embedded into a larger unitary as:
\begin{footnotesize}
\begin{align}
\frac{\mathbf{H}}{\lambda}=-\begin{pmatrix}
\mathbf{0}_{N^2} & (\mathbb{I}_N\otimes \ket{0}\bra{0}^{\otimes n}\otimes \bra{0}^{\otimes r+2})\mathcal{U}_{\mathbf{B}}(\mathbb{I}_N\otimes\mathbb{I}_N\otimes\ket{0}^{\otimes r+2})\\
(\mathbb{I}_N\otimes\mathbb{I}_N\otimes\bra{0}^{\otimes r+2})\mathcal{U}_{\mathbf{B}}^\dagger(\mathbb{I}_N\otimes \ket{0}\bra{0}^{\otimes n}\otimes \ket{0}^{\otimes r+2}) & \mathbf{0}_{N^2}
\end{pmatrix}\label{Hamiltonian2}
\end{align} 
\end{footnotesize}
Here, the Hamiltonian is of size $2N^2 \times 2N^2$, acting on a quantum register of $2n + 1$ qubits. The unitary $\mathcal{U}_{\mathbf{B}}$ operates on $2n + r + 1$ qubits, while the block encoding in Eq.~\eqref{Hamiltonian2} requires two additional ancilla qubits:
(1) an ancilla to select between the upper-right and lower-left blocks of the Hamiltonian, and
(2) an ancilla to implement a block encoding of the projector $\mathbf{P}_{N^2} := \ket{0}\bra{0}^{\otimes n}$, which is non-unitary by nature.
The resulting quantum circuit structure is illustrated in Fig.~\ref{fig:BEH_circuit}.


To clarify the action of this circuit, consider the input state
\begin{equation}
    \ket{\Phi}=\frac{1}{\sqrt{2}}\ket{0}_{\mathbf{P}}\otimes\left(\ket{0}_{\mathbf{U}}\otimes\ket{\phi_v}_\text{T}+\ket{1}_{\mathbf{U}}\otimes\ket{\phi_x}_\text{T}\right)
\end{equation}
where $\ket{0}{\mathbf{U}} \otimes \ket{\phi_v}\text{T}$ corresponds to the velocity component (first line of Eq.~\eqref{eq:subnorm}),
and $\ket{1}{\mathbf{U}} \otimes \ket{\phi_x}\text{T}$ corresponds to the position terms (second and third lines of Eq.~\eqref{eq:subnorm}, up to ancilla).

The full protocol consists of the following controlled operations:
\begin{enumerate}
\item Hadamard gates on $\ket{0}_{\mathbf{P}}$:
\begin{align*}
\ket{\Phi_1}=\frac{1}{\sqrt{2}}\left[\right.&+\ket{0}_{\mathbf{P}}\otimes \ket{0}_{\mathbf{U}}\otimes \ket{\phi_v}_\text{T}\\
&+\left.\ket{0}_{\mathbf{P}}\otimes \ket{1}_{\mathbf{U}}\otimes \ket{\phi_x}_\text{T}\right.\\
&+\left.\ket{1}_{\mathbf{P}}\otimes \ket{0}_{\mathbf{U}}\otimes \ket{\phi_v}_\text{T}\right.\\
&+\left.\ket{1}_{\mathbf{P}}\otimes \ket{1}_{\mathbf{U}}\otimes \ket{\phi_x}_\text{T}\right]. 
\end{align*}
\item Control on ancilla $\mathbf{U}$, acting $-\mathbf{U}_{\mathbf{B}}$ on $\ket{\phi_v}_\text{T}$:
\begin{align*}
\ket{\Phi_2}=\frac{1}{\sqrt{2}}\left[\right.&-\left.\ket{0}_{\mathbf{P}}\otimes \ket{0}_{\mathbf{U}}\otimes  (\mathbf{U}_{\mathbf{B}}\ket{\phi_v}_\text{T})\right.\\
&+\left.\ket{0}_{\mathbf{P}}\otimes \ket{1}_{\mathbf{U}}\otimes \ket{\phi_x}_\text{T}\right.\\
&-\left.\ket{1}_{\mathbf{P}}\otimes \ket{0}_{\mathbf{U}}\otimes (\mathbf{U}_{\mathbf{B}}\ket{\phi_v}_\text{T})\right.\\
&+\left.\ket{1}_{\mathbf{P}}\otimes \ket{1}_{\mathbf{U}}\otimes \ket{\phi_x}_\text{T}\right]. 
\end{align*}
\item Control on ancilla $\mathbf{P}$, acting $2\mathbf{P}_N-\mathbf{I}_N$ on $\ket{\phi_v}_\text{T}$:
\begin{align*}
\ket{\Phi_3}=\frac{1}{\sqrt{2}}\left[\right.&-\left.\ket{0}_{\mathbf{P}}\otimes \ket{0}_{\mathbf{U}}\otimes ((2\mathbf{P}_N-\mathbf{I}_N)\mathbf{U}_{\mathbf{B}}\ket{\phi_v}_\text{T})\right.\\
&+\left.\ket{0}_{\mathbf{P}}\otimes \ket{1}_{\mathbf{U}}\otimes \ket{\phi_x}_\text{T}\right.\\
&-\left.\ket{1}_{\mathbf{P}}\otimes \ket{0}_{\mathbf{U}}\otimes (\mathbf{U}_{\mathbf{B}}\ket{\phi_v}_\text{T})\right.\\
&+\left.\ket{1}_{\mathbf{P}}\otimes \ket{1}_{\mathbf{U}}\otimes \ket{\phi_x}_\text{T}\right]. 
\end{align*}
\item Control on ancilla $\mathbf{P}$, acting $2\mathbf{P}_N-\mathbf{I}_N$ on $\ket{\phi_x}_\text{T}$:
\begin{align*}
\ket{\Phi_4}=\frac{1}{\sqrt{2}}\left[\right.&-\left.\ket{0}_{\mathbf{P}}\otimes \ket{0}_{\mathbf{U}}\otimes ((2\mathbf{P}_N-\mathbf{I}_N)\mathbf{U}_{\mathbf{B}}\ket{\phi_v}_\text{T})\right.\\
&+\left.\ket{0}_{\mathbf{P}}\otimes \ket{1}_{\mathbf{U}}\otimes ((2\mathbf{P}_N-\mathbf{I}_N))\ket{\phi_x}_\text{T})\right.\\
&-\left.\ket{1}_{\mathbf{P}}\otimes \ket{0}_{\mathbf{U}}\otimes (\mathbf{U}_{\mathbf{B}}\ket{\phi_v}_\text{T})\right.\\
&+\left.\ket{1}_{\mathbf{P}}\otimes \ket{1}_{\mathbf{U}}\otimes \ket{\phi_x}_\text{T}\right]. 
\end{align*}
\item Control on ancilla $\mathbf{U}$, acting $-\mathbf{U}_B^\dagger$ on $\ket{\phi_x}_\text{T}$:
\begin{align*}
\ket{\Phi_5}=\frac{1}{\sqrt{2}}\left[\right.&-\left.\ket{0}_{\mathbf{P}}\otimes \ket{0}_{\mathbf{U}}\otimes ((2\mathbf{P}_N-\mathbf{I}_N)\mathbf{U}_{\mathbf{B}}\ket{\phi_v}_\text{T})\right.\\
&-\left.\ket{0}_{\mathbf{P}}\otimes \ket{1}_{\mathbf{U}}\otimes (\mathbf{U}_{\mathbf{B}}^\dagger(2\mathbf{P}_N-\mathbf{I}_N))\ket{\phi_x}_\text{T})\right.\\
&-\left.\ket{1}_{\mathbf{P}}\otimes \ket{0}_{\mathbf{U}}\otimes (\mathbf{U}_{\mathbf{B}}\ket{\phi_v}_\text{T})\right.\\
&-\left.\ket{1}_{\mathbf{P}}\otimes \ket{1}_{\mathbf{U}}\otimes (\mathbf{U}_{\mathbf{B}}^\dagger\ket{\phi_x}_\text{T})\right]. 
\end{align*}
\item Applying Hadamard to the ancilla register $\mathbf{P}$:
\begin{align*}
\ket{\Phi_\text{f}}=-\ket{0}_\mathbf{P}\otimes\left[\right.&\ket{0}_\mathbf{U}\otimes\left(\mathbf{P}_N\mathbf{U}_\mathbf{B}\ket{\phi_v}_\text{T}\right)\\+&\left.\ket{1}_\mathbf{U}\otimes\left(\mathbf{U}^\dagger_\mathbf{B}\mathbf{P}_N\ket{\phi_x}_\text{T}\right)\right]\\
-\ket{1}_\mathbf{P}\otimes\left[\right.&\ket{0}_\mathbf{U}\otimes\left((\mathbf{P}_N-\mathbf{I}_N)\mathbf{U}_\mathbf{B}\ket{\phi_v}_\text{T}\right)\\+&\left.\ket{1}_\mathbf{U}\otimes\left(\mathbf{U}^\dagger_\mathbf{B}(\mathbf{P}_N-\mathbf{I}_N)\ket{\phi_x}_\text{T}\right)\right].
\end{align*}
\end{enumerate}

This final state reflects the action of a block-encoded Hamiltonian on the input state $\ket{\Phi}$, with the desired Hamiltonian action contained in the subspace where the ancilla register $\mathbf{P}$ corresponds to the state $\ket{0}_\mathbf{P}$.


\end{document}